\documentclass[fleqn,usenatbib]{mnras}

\usepackage[T1]{fontenc}
\usepackage{ae,aecompl}

\usepackage{graphicx}	
\usepackage{amsmath}	
\usepackage{amssymb}	

\usepackage{multirow}
\renewcommand{\[}{\begin{equation}}
\renewcommand{\]}{\end{equation}}

\def\Gyr{\,\mathrm{Gyr}}
\def\Myr{\,\mathrm{Myr}}
\def\kpc{\,\mathrm{kpc}}

\def\kms{\,\mathrm{km\,s}^{-1}}

\def\msun{\,{\rm M}_\odot}

\def\vlos{{V_\parallel}}

\newcommand{\gbp}{{G_\mathrm{BP}}}
\newcommand{\grp}{{G_\mathrm{RP}}}

\newcommand{\vl}{V_\ell}
\newcommand{\vb}{V_b}

\newcommand{\vphi}{V_\phi^*}
\newcommand{\vz}{V_Z^*}
\newcommand{\lz}{L_Z^*}
\newcommand{\rg}{R_g^*}
\newcommand{\vphireal}{V_\phi}
\newcommand{\vzreal}{V_Z}
\newcommand{\vR}{V_R}

\newcommand{\usun}{U_\odot}
\newcommand{\vsun}{V_\odot}
\newcommand{\wsun}{W_\odot}
\newcommand{\rsun}{R_\odot}
\newcommand{\zsun}{Z_\odot}

\newcommand\gaia{\textit{Gaia}}

\newcommand\gdrtwo{\gaia~DR2}
\newcommand\egdr[1]{\gaia~EDR#1}

\newcommand{\agama}{\textsc{agama}}

\newcommand{\orcidicon}[1]{\href{https://orcid.org/#1}{\includegraphics[height=\fontcharht\font`\B]{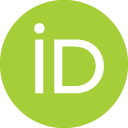}}}

\def\referee{ }

\usepackage{newtxtext,newtxmath}

\defcitealias{AnticentrePaper}{AC21}

\title[The disturbed outer Galaxy]{The disturbed outer Milky Way disc}

\author[P. J. McMillan et al.]{Paul J. McMillan$^{1}$\thanks{E-mail: paul@astro.lu.se} \orcidicon{0000-0002-8861-2620},
Jonathan Petersson$^{1}$ \orcidicon{0000-0001-7248-3898},
Thor Tepper-Garcia$^{2,3}$
\orcidicon{0000-0002-1081-883X},\newauthor
Joss Bland-Hawthorn$^{2,3}$
\orcidicon{0000-0001-7516-4016},
Teresa Antoja$^{4,5,6}$ \orcidicon{0000-0003-2595-5148},
Laurent Chemin$^7$ \orcidicon{0000-0002-3834-7937},\newauthor
Francesca Figueras$^{4,5,6}$ \orcidicon{0000-0002-3393-0007},
Shourya Khanna$^8$ \orcidicon{0000-0002-2604-4277},
Georges Kordopatis$^9$ \orcidicon{0000-0002-9035-3920}, \newauthor
Pau Ramos$^{4,5,6}$ \orcidicon{0000-0002-5080-7027}, 
Merce Romero-G\'omez$^{4,5,6}$ \orcidicon{0000-0003-3936-1025},
George Seabroke$^{10}$ \orcidicon{0000-0003-4072-9536}
\\
$^{1}$Lund Observatory, Department of Astronomy and Theoretical Physics, Lund University, Box 43, SE-22100, Lund, Sweden\\
$^{2}$Sydney Institute for Astronomy, School of Physics, University of Sydney, NSW 2006, Australia\\
$^{3}$Centre of Excellence for All Sky Astrophysics in Three Dimensions (ASTRO-3D), Australia \\
$^{4}${Departament de Física Qu\`antica i Astrof\'isica (FQA), Universitat de Barcelona (UB),  c. Mart\'i i Franqu\`es, 1, 08028 Barcelona, Spain}\\
$^{5}${Institut de Ci\`encies del Cosmos (ICCUB), Universitat de Barcelona (UB), c. Mart\'i i Franqu\`es, 1, 08028 Barcelona, Spain}\\
$^{6}${Institut d'Estudis Espacials de Catalunya (IEEC), c. Gran Capit\`a, 2-4, 08034 Barcelona, Spain} \\
$^{7}$Centro de Astronom\'ia, Universidad de Antofagasta, Avda. U. de Antofagasta 02800, Antofagasta, Chile \\
$^{8}$INAF - Osservatorio Astrofisico di Torino, via Osservatorio 20, 10025 Pino Torinese (TO), Italy \\
$^{9}$Universit\'{e} C\^{o}te d'Azur, Observatoire de la C\^{o}te d'Azur, CNRS, Laboratoire Lagrange, Bd de l'Observatoire, CS 34229, 06304 Nice Cedex 4, France\\
$^{10}$Mullard Space Science Laboratory, University College London, Holmbury St Mary, Dorking, Surrey, RH5 6NT, United Kingdom.\\
}

\pubyear{2022}

\begin{document}
\label{firstpage}
\pagerange{\pageref{firstpage}--\pageref{lastpage}}
\maketitle

\begin{abstract}
The outer parts of the Milky Way's disc are significantly out of equilibrium. 
Using only distances and proper motions of stars from \gaia's Early Data Release 3, in the range $|b|<10^\circ$, $130^\circ<\ell<230^\circ$, we show that for stars in the disc between around $10$ and $14\kpc$ from the Galactic centre, vertical velocity is strongly dependent on the angular momentum, azimuth, and position above or below the Galactic plane. We further show how this behaviour translates into a bimodality in the velocity distribution of stars in the outer Milky Way disc.
We use an $N$-body model of an impulse-like interaction of the Milky Way disc with a perturber similar to the Sagittarius dwarf to demonstrate that this mechanism can generate a similar disturbance. It has already been shown that this interaction can produce a phase spiral similar to that seen in the Solar neighbourhood. We argue that the details of this substructure in the outer galaxy will be highly sensitive to the timing of the perturbation or the gravitational potential of the Galaxy, and therefore may be key to disentangling the history and structure of the Milky Way.
\end{abstract}




\section{Introduction}

The European Space Agency's {\gaia} mission \citep{Prusti2016} has allowed us to see with new clarity the extent to which the Milky Way's disc is out of equilibrium. The data from {\gaia} includes photometry and spectra, but it is the extraordinarily precise astrometry that sets Gaia furthest apart from any other instrument.

With the releases of {\gdrtwo} and EDR3 \citep{Brown2018,EDR3}, new disturbances have been discovered, and ones that were already known have been mapped in far greater detail. Substructure in velocities parallel to the disc have been mapped locally and across the disc  
\citep[e.g.,][]{Dehnen1998,Famaey2005,Antoja2008,Katz2018,Kawata2018,Trick2019,Friske2019,Khanna2022}. Vertical asymmetries have been found in both number counts and velocities, and associated with both the Galactic warp and with bending or breathing modes in the disc  \citep[e.g.][]{Widrow2012,Williams2013,Schonrich2018,Poggio2018,Bennett2019,Carrillo2019,Khanna2019b,Cheng2020}. 

The Gaia phase spiral\footnote{There is no consensus in the current literature on how we should refer to this feature, which was originally dubbed the `snail shell'. While we would prefer `kanelbulle', reflecting the similarity with the shape of the Swedish cinnamon roll, we choose `phase spiral' in this study to follow the most common choice.} 
was discovered in {\gdrtwo} data by \cite{Antoja2018} and is an overdensity of stars along a spiral in the $Z-V_Z$ plane, where $Z$ is the direction perpendicular to the Galactic disc. It is also found as a spiral shaped variation in the average $\vR$ or $V_\phi$ velocities in the $Z-V_Z$ plane.  A number of studies have demonstrated that this can be produced as a perturbation induced by the passage of the Sagittarius (Sgr) dwarf galaxy phase-mixes over time \citep{Binney2018,Darling2019,Laporte2019,Bland-Hawthorn2021,2022ApJ...928...80G}, and this is now the generally favoured formation scenario. Other studies have investigated how this structure persists and varies across the galaxy as a function of chemistry, position and kinematics 
\citep[e.g.,][Alinder, McMillan \& Bensby in prep.]{Bland-Hawthorn2019,Xu.2020,Li2021}. 

In connection with {\egdr3}, \citet[henceforth \citetalias{AnticentrePaper}]{AnticentrePaper} looked at stars in the Galactic anticentre ($170^\circ<\ell<190^\circ$, $|b|<10^\circ$), and showed that in this region the velocity distribution in the outer disc was significantly disturbed. At Galactic radii between $10$ and $14\kpc$  stars predominantly follow a bimodal velocity distribution. One of these modes is at $V_Z\approx10\kms$ and is mostly populated by stars below the Galactic plane, while the other is at $V_Z\approx-5\kms$, is mostly populated by stars above the Galactic plane, and lags the rotation of the first group by around $30\kms$. We, perhaps inelegantly, refer to these two modes as `clumps' throughout this study. The strength of the two clumps varies strongly with increasing radius, with the clump at positive $V_Z$ becoming increasingly dominant at greater radii.

This behaviour is elegantly, if incompletely, summarised by plotting the distribution in $V_Z$ velocity of stars at different angular momenta. \citetalias{AnticentrePaper} showed that towards the Galactic anticentre this distribution has a clear break from negative $V_Z$ to positive at an angular momentum around $|L_Z|=2750 \kms\kpc$, corresponding to a guiding centre radius of ${\sim}11.5\kpc$.


The \citetalias{AnticentrePaper} study was limited to the anticentre region because in this region the velocity in the plane of the sky closely corresponds to the Galactocentric azimuthal and vertical velocities. This allowed the {\gaia} team to gain significant insight into the velocity structure of the outer disc of the Milky Way using a sample of stars without measured line-of-sight velocities\footnote{Throughout this text we will refer to `line-of-sight' velocities rather than `radial' velocities to avoid confusion with the velocity component in the Galactocentric radial direction.}. The productive use of proper motion samples alone to study Galactic dynamics has a long history, including the work by Eggen on moving groups \citep[e.g.,][]{Eggen1958}; the characterisation of the local velocity distribution by \cite{DehnenBinney1998}, \cite{Dehnen1998} and recently for the white dwarf population by \cite{Mikkola2022}; measurement of velocity asymmetry in the Galactic disc \citep{Antoja2017}; and the determination of the Milky Way's escape velocity by \cite{Koppelman2021}. 

In this study we extend the approach of \citetalias{AnticentrePaper} to include stars over a wide area in the outer disc. Because we are not limited to stars with measured line-of-sight velocities, we can use a far larger sample of stars than would otherwise be available. This allows us to explore the region with a level of detail that is not otherwise possible. We investigate whether the bimodality of the velocity distribution is only found around the Galactic anticentre, or if it can be found elsewhere in the galaxy. We ask whether we can see any change in it as we move around the galaxy.

The paper is organised as follows: In Section~\ref{sec:data} we describe the dataset we study and the approximations we make to allow us to draw conclusions from these data. We also demonstrate the validity of these approximations on mock data. In Section~\ref{sec:OuterMilkyWay} we look at the disturbances in this part of the outer Milky Way disc. In Section~\ref{sec:simulations} we present $N$-body simulations of an impulse-like interaction with the Milky Way disc, and show that it can produce similar behaviour in the outer disc to that which we see. Finally we discuss  and summarise our results in Secs.~\ref{sec:discuss}~\&~\ref{sec:conclusions}, respectively.

\section{Data} \label{sec:data}

\begin{figure}
\centering
	\includegraphics[width=0.8\hsize]{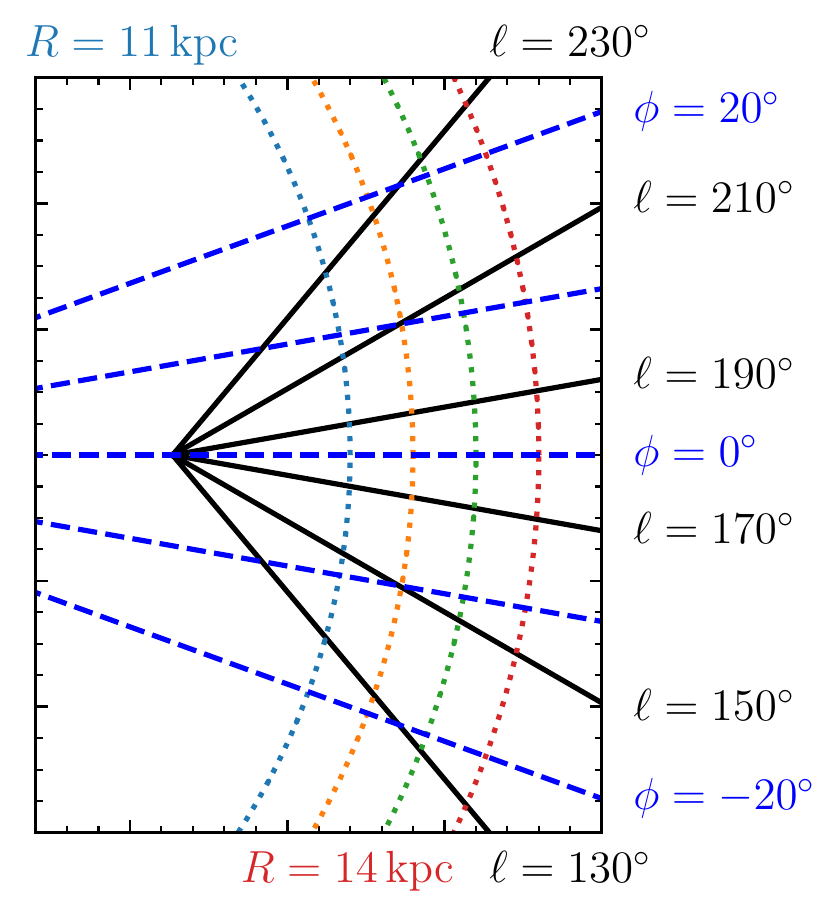}
    \caption{Illustration of the range in Galactocentric $\phi$ angles covered by ranges in $\ell$ at different points in the Galaxy (in the area we consider). Our {\gaia} data come from $130^\circ<\ell<230^\circ$, while in Sec~\ref{sec:simulations} we look at particles in the simulation in the range $-20^\circ<\phi<20^\circ$. Solid black line are lines of constant $\ell$, and converge at the position of the Sun. Blue dashed lines are lines of constant $\phi$. Lines of constant Galactocentric radius are shown as dotted lines at $R=11,12,13$ \& $14\kpc$ coloured, respectively, light-blue, orange, green and red. } 
    \label{fig:PhiVsl}
\end{figure}

In this work we investigate the dynamics of stars in the outer galaxy, with Galactic longitudes in the range $130^\circ<\ell<230^\circ$ and latitudes $|b|<10^\circ$.  The range of Galactocentric azimuths covered by these data are illustrated in Fig.~\ref{fig:PhiVsl}. The data are taken from {\egdr3} and consist of the proper motions $\mu_\alpha*$, $\mu_\delta$; and the photometric magnitudes $G$, $\gbp$, $\grp$. We take the distance to the stars to be the photogeometric estimates from \cite{BailerJones21}. These estimates take into account the observed colours and magnitudes of stars, along with their parallaxes \citep[corrected following the recipe from][]{2021A&A...649A...4L} and a prior constructed from a three-dimensional model of our Galaxy, to derive distance estimates that can be significantly more precise than those derived without photometry.

Importantly, we do not restrict ourselves to only those stars with measured line-of-sight velocities. {\egdr3} contains over 1.4 billion stars with astrometric measurements, of which only 7 million (i.e., less than 0.5 percent) have published {\gaia} line-of-sight velocities. After \gaia\ DR3 in June 2022 this number will rise, but only to around 33 million (i.e., around 2.5 percent of the full catalogue). Our choice not to use them, therefore, enables us to work with a very large sample of stars, but at the cost of losing one component of the velocity. As we shall demonstrate, we are able to make approximations that allow us to study the velocity distribution in the vertical and azimuthal directions with just these data. However these approximations become less reasonable away from $b=0^\circ$ and $\ell=180^\circ$. 

To ensure that we are using high quality data we require that the astrometric `Renormalised Unit Weight Error' ({\tt RUWE}) meets the criterion $\mathtt{RUWE}<1.4$. We also require that the parallax divided by its error $\varpi/\sigma_\varpi>3$, so that the photogeometric distance estimation is not completely dominated by the photometric information. 
With these criteria, we have $20\,041\,385$ stars in our full sample.

\begin{figure}
	\includegraphics[width=\hsize]{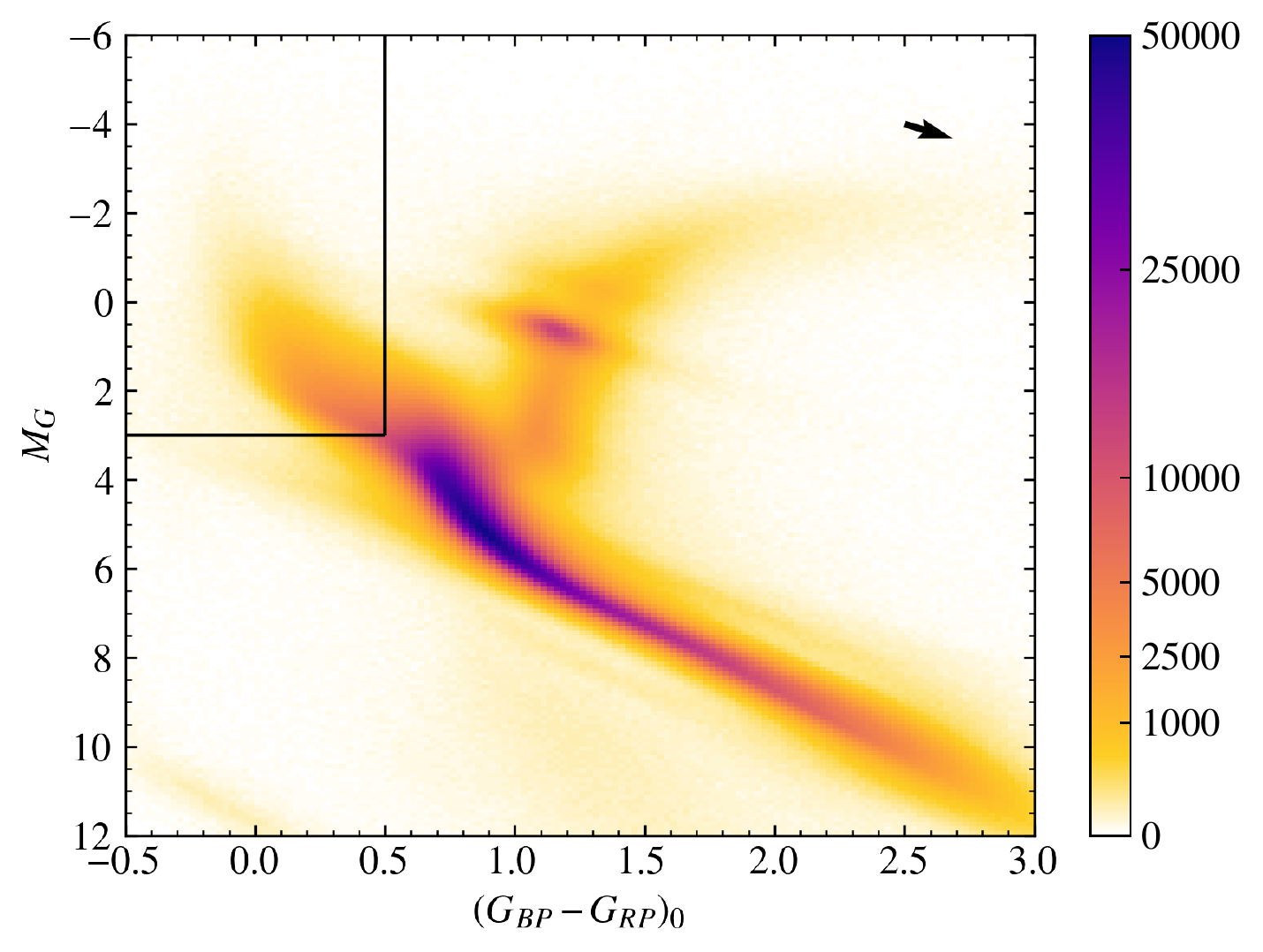}
    \caption{Extinction-corrected colour-magnitude diagram for stars. The box at the top left contains the stars in our young blue main-sequence sub-sample. The quoted numbers are numbers of stars in a pixel of width $0.025$ mag in colour and height $0.1$ mag in absolute magnitude. The arrow shown at the top right corner of the diagram illustrates the assumed reddening vector. Stars will be moved in this direction on the diagram depending on whether the extinction to them has been over- or under-estimated. } 
    \label{fig:HRdiagram}
\end{figure}

\cite{Bland-Hawthorn2019} argued that the kinematically colder populations in the Milky  Way disc are more strongly affected by the phase spiral than kinematically warmer populations. Further, several authors \citep[e.g.,][]{Amores2017,Poggio2018,RomeroGomez19} have shown that the structure of the Galactic warp depends on the age of the tracer population considered. We therefore construct a sub-sample dominated by young blue main-sequence stars as indicated in Fig~\ref{fig:HRdiagram}. Extinction is taken as the median extinction at the quoted distance in the dust map of \cite{Green2019}, and the conversion from $E(B-V)$ to the {\gaia} bands is approximated as $A_G = 2.294\, E(B-V)$; $E(G_{BP}-G_{RP}) = 1.309\,E(B-V)$, figures we have taken from  \cite{Sanders2018}.

We define our blue sample with a simple cut in extinction-corrected absolute magnitude and colour
\begin{align}
    M_G < 3.0 \\
    (G_{BP}-G_{RP})_{\,0} < 0.5 
\end{align}
as illustrated in Fig.\ref{fig:HRdiagram}. This leaves us with a sample of $1\,093\,192$ stars.

The selection of these stars is advantageous for two further reasons: Firstly, because the stars are relatively bright, their \gaia\ parallax measurements will have relatively small uncertainties; secondly, because their velocity dispersion in the Galactocentric radial direction is typically very small, the uncertainty due to the absence of a measured line-of-sight velocity is reduced (see below). 

\subsection{Velocity estimates} \label{sec:techniques}

 If we have a measured distance and proper motion of a star, then we know two components of the star's velocity, which we can decompose into the direction of increasing Galactic coordinates $\ell$ and $b$ (we refer to these as $\vl$ and $\vb$ respectively). Without a measured line-of-sight velocity for a star we cannot know its full 3D velocity.
 
\begin{figure*}
	\includegraphics[width=\hsize]{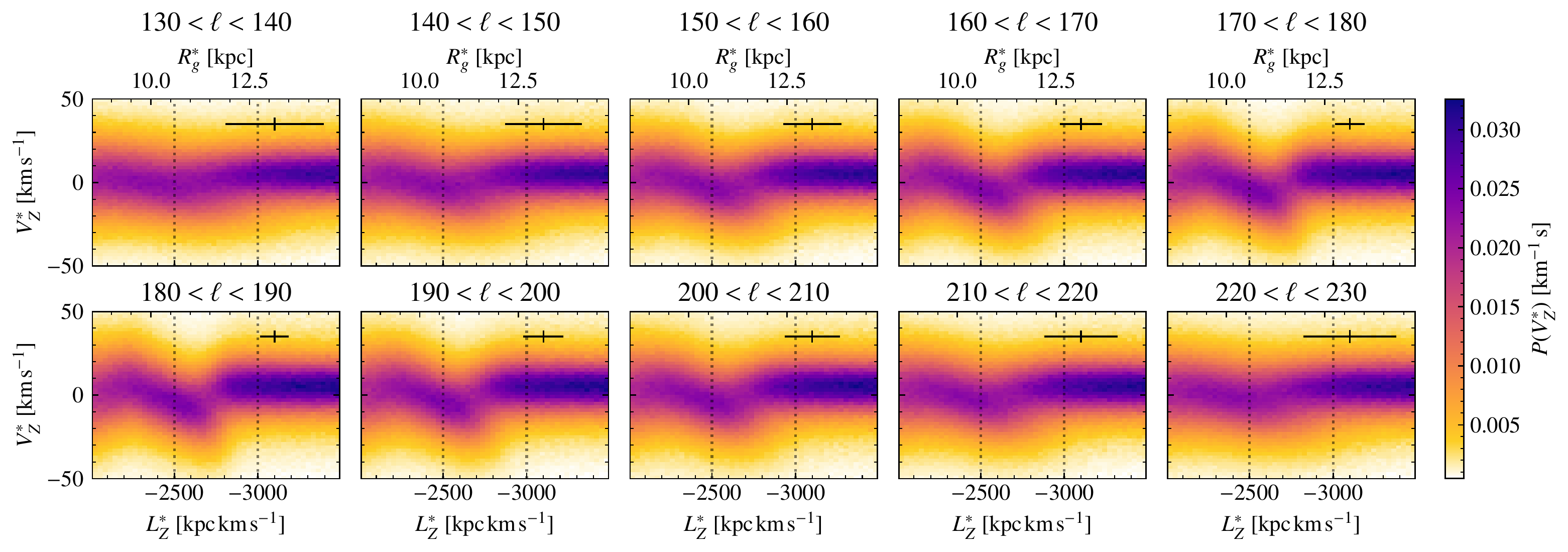}
    \caption{Density in $\vz$ at different $\lz$ for mock data with distance uncertainties and no line-of-sight velocity measurement. Each panel shows the results for different $10^\circ$ ranges in $\ell$. Vertical dotted lines are drawn at $\lz={-}2500$~\&~${-}3000\kms\kpc$ to guide the eye and to ease the comparison between equivalent plots elsewhere in the text. Error bars in the top right of each panel indicate the typical uncertainties in $\lz$ and $\vz$. The underlying pattern is seen relatively clearly in panels near $\ell=180^\circ$, but becomes more smeared out further from the anticentre.} 
    \label{fig:FullySmeared}
\end{figure*}

\begin{figure}
	\centering\includegraphics[width=7cm]{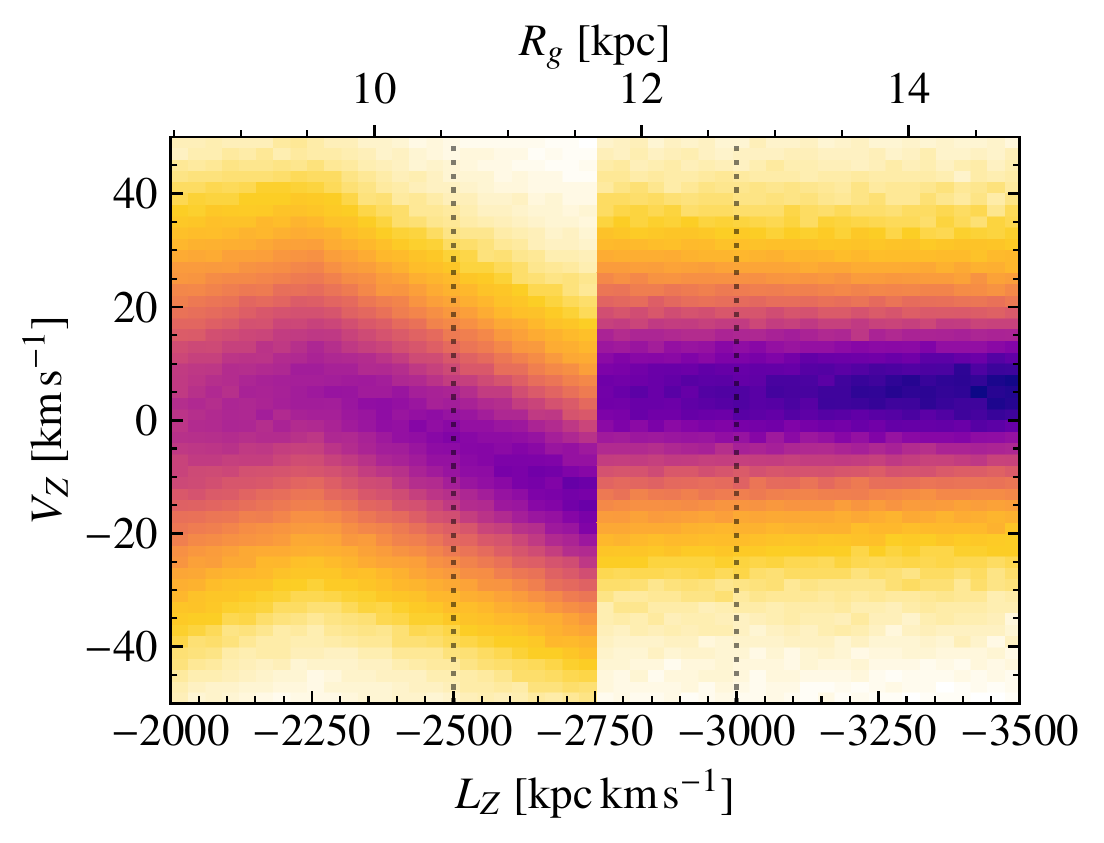}
    \caption{True density in $\vzreal$ at each $L_Z$ for our mock data. Dotted lines are drawn at $\lz={-}2500$~\&~${-}3000\kms\kpc$ to guide the eye. The model is based on one in equilibrium (which would have average $\vzreal=0$ for all $L_z$), perturbed as described in the text to mimic the structure seen in the Milky Way. } 
    \label{fig:Model}
\end{figure}
 
To understand motions in a galaxy it is convenient to consider velocities in terms of their Galactocentric components in cylindrical polar coordinates $(\vR,\vzreal,\vphireal)$. The relationship between $(\vl,\vb)$ and $(\vR,\vzreal,\vphireal)$ depends on a star's position in the Galaxy, and towards the Galactic anticentre $(\ell,b=180^\circ,0^\circ)$ and centre $(\ell,b=0^\circ,0^\circ)$, we have $\vl = \pm \vphireal$ and $\vb=\vzreal$, while proper motions provide no information about $\vR$. This convenience has led a number of authors to focus studies in these regions \citep[e.g.,][\citetalias{AnticentrePaper}]{Kawata2018,FernandezAlvar2021}.

At other Galactic coordinates we do not have this convenient relationship. However, we can make further progress by making a simple approximation: That the typical $\vR\approx0$ at any point in the disc. There is a dispersion around this value, which will produce a corresponding uncertainty in the derived values which increases as we move further from $\ell=180^\circ$.
We know that even the average $\vR$  is not exactly zero across the Galaxy, with \gdrtwo\  in particular demonstrating that it varies at the level of ${\sim}10\kms$ 
{\referee in the extended Solar neighbourhood} \citep{Katz2018}, {\referee while \cite{Eilers2020} showed that it varies with an amplitude of ${\sim7\kms}$ in the outer disc}.
However, this is much smaller than the typical $\vphireal$ velocity in the disc, so makes a fairly small difference. We quantify the errors introduced by this approximation in Section \ref{sec:CheckvR}.

For a star in any position in the Galaxy with measured $(\vl,\vb)$, there is a  line-of-sight velocity, which we call $\vlos^*$, that would give it $\vR=0$. We can show that this velocity
\begin{equation}
    \vlos^* = \frac{X v_X^\bot +Y v_Y^\bot}{(X\cos l + Y\sin l)\cos b},
\end{equation}
where $(X,Y,Z)$ are Cartesian components of a star's position in the Galaxy\footnote{This coordinate system has the Galactic centre at $(0,0,0)$, and the Sun at $(\rsun,0,\zsun)$. $(R,\phi,Z)$ are the cylindrical coordinates associated with this coordinate system.}, and $(v_X^\bot,v_Y^\bot,v_Z^\bot)$ are the components of the star's Galactocentric velocity in the $X,Y,Z$ coordinate system that come only from its velocity in the plane of the sky. We can use this to find our estimates for $\vphireal$ and $\vzreal$, which are
\begin{equation}
    \vphi = \frac{R(-v_X^\bot\sin \ell + v_Y^\bot\cos \ell)}{X\cos \ell + Y \sin \ell}
\end{equation}
\begin{equation}
    \vz = v_Z^\bot + \vlos^* \sin b
\end{equation}
A full derivation of these equations is in Appendix \ref{app:Convert}.

{\referee One could consider taking more sophisticated approaches than the approximation that $\vR=0$. The value of $\vR$ could be estimated as a function of position using the stars for which there are measurements of the line-of-sight velocity. This would have to be done carefully to avoid adding excess noise, especially far from the Sun where there are few such stars. It would also be taking its $\vR$ estimate from a different (brighter) population than the population being considered. Recently, it has been proposed that a Bayesian neural network can be used to estimate the missing line-of-sight velocities for Gaia stars \citep{2022arXiv220604102N}, but it is not clear whether this provides a better point estimate than our approach. In this study our simple approach is sufficient, and has the benefit of being straightforwardly reproducible and less susceptible to unexpected systematic errors. We defer the study of improvements we could make for this approximation to future work. }

We have to assume a position and velocity for the Sun in the Milky Way to convert our measurements into the Galactocentric frame. We assume the Sun is located a distance  $R_\odot=8.178\kpc$ from the Galactic centre \citep{Gravity2019} and a height above the Galactic plane of $Z_\odot=0.02\kpc$ \citep{Bennett2019}. We further assume $(v_{X,\odot},v_{Y,\odot}, v_{Z,\odot}=(-11.1, -248.5, 7.25)\kms$ for the solar motion \citep*{Schonrich2010,Reid2020}.

Throughout this paper we indicate that a quantity has been estimated without a line-of-sight velocity measurement by giving it an asterisk. Our estimate of the angular momentum of a star is $\lz\equiv R\vphi$, which for most stars in the Milky Way disc is negative. In the interests of making our results easier to interpret, we augment this throughout this article with a simple estimate of the guiding centre radius $\rg=|\lz|/(236\kms)$.


\subsection{Demonstration on mock data} \label{sec:CheckvR}
To test our assumption that we can proceed using only proper motions, we apply our method to mock data with known properties. We base this on a dynamical model constructed using the \agama\ software package \citep{Vasiliev2019}. 
This model consists of two disc components and a stellar halo, and is designed to be in equilibrium within a given axisymmetric gravitational potential \citep[the main model from][]{McMillan17}. We sample $40\,000\,000$ particles from the model which, since the model is axisymmetric, we can place at any position angle in the Galaxy we are interested in. The samples we construct from this model, which come from $10^\circ$ ranges in $\ell$ between $130^\circ$ and $230^\circ$, contain between $2\,800\,000$ and $3\,200\,000$ particles when we apply the constraint that they also feature only stars with $|b|<10^\circ$.

\begin{figure*}
	\includegraphics[width=\hsize]{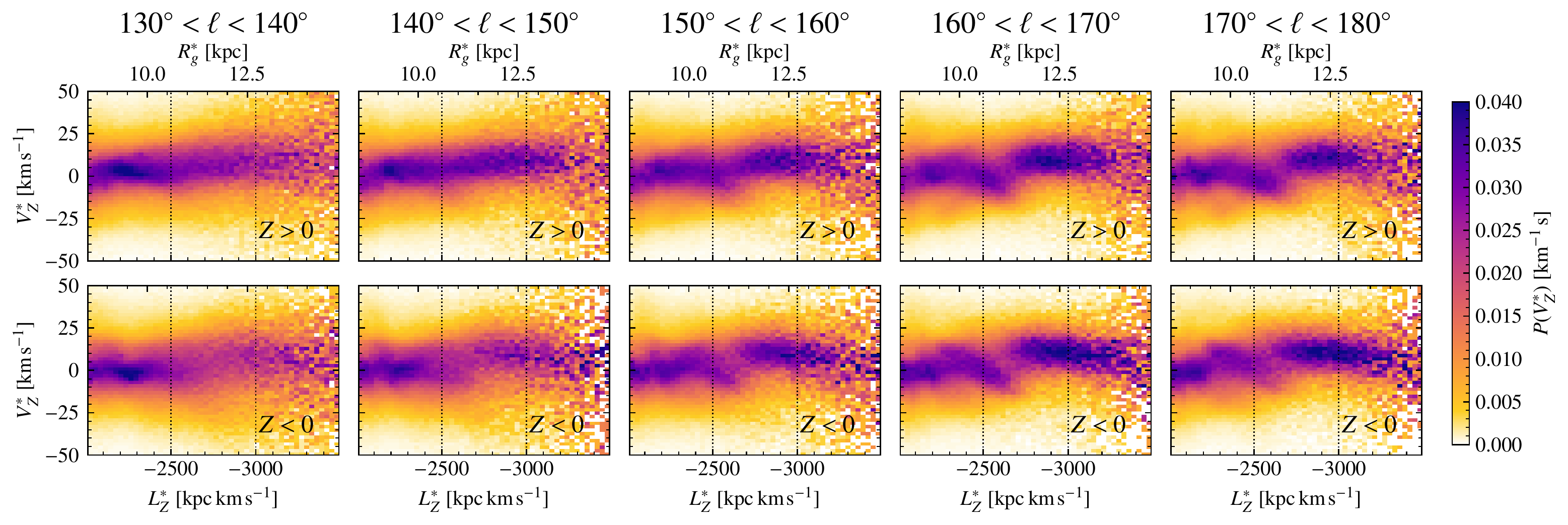}
	\includegraphics[width=\hsize]{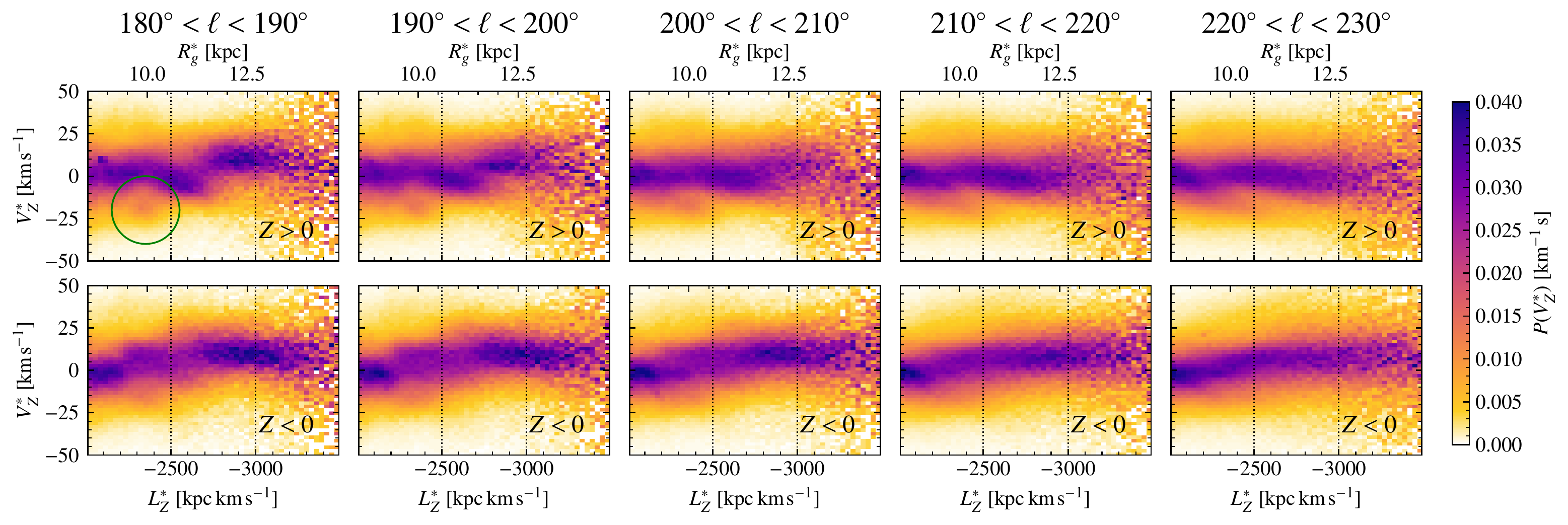}
    \caption{The distribution in $\vz$ of stars in the {\gaia} data as a function of their estimated angular momentum $\lz$ or, equivalently, guiding centre radius $\rg$. These are shown for $10^\circ$ ranges in $\ell$, ranging from $130^\circ$ to $180^\circ$ for the upper set of panels and $180^\circ$ to $230^\circ$ for the lower set of panels. In each case this is also divided into the distribution above and below the plane. Vertical dotted lines are again drawn at $\lz={-}2500$~\&~${-}3000\kms\kpc$ to guide the eye and to ease the comparison between equivalent plots elsewhere in the text. The green circle in the panel for $Z>0$ at $180^\circ< l<190^\circ$ highlights a feature which can be seen in a number of panels. The `break' seen in the anticentre region at $\lz\approx2750\kms\kpc$ is clearly visible in many panels, with noticeable differences above and below the plane.} 
    \label{fig:RgVz}
\end{figure*}

We perturb the vertical velocities of stars in this model to (very approximately) mimic the structure seen by \citetalias{AnticentrePaper}. We change the vertical velocity of each star by an amount $\Delta\vzreal\kms$ that is a function of the $Z$ component of angular momentum, $L_z$. For convenience we express this in terms of $l_Z=|L_Z|/(\kms\kpc)$:
\begin{equation}
    \Delta\vzreal = \begin{cases}
    0 & \text{for }l_Z \leq 2000 \\
    6\left(l_Z-2000\right)/250 & \text{for }2000 < l_Z \leq 2250 \\
    6-20\left(l_Z-2250\right)/500 & \text{for }2250 < l_Z \leq 2750  \\
    5 & \text{for }2750 < l_Z.
    \end{cases}
\end{equation}

To demonstrate that our method is robust against the variations in $\vR$ in the Galaxy, we further perturb the model with a term that mimics the influence of the spiral arms on the velocity field of the outer Milky Way,{\referee guided by the results from \cite{Eilers2020}}. We assume that the influence of the spiral arms is primarily on  $\vR$, and perturb the velocities of all stars in our model by an amount $\Delta \vR$ corresponding to an $m=2$ spiral perturbation with
\begin{equation}
    \Delta \vR(R,\phi) = V_{\rm max} \sin\left[m\left(\phi - \phi_0 - \frac{\log(R/\kpc)}{\tan(p)}\right)\right], 
\end{equation}
where we take amplitude $V_{\rm max}=7\kms$, $\phi_0=0^\circ$ and pitch angle $p=12^\circ$. This produces a spiral perturbation that is comparable to that found in the outer disc by \cite{Eilers2020}, or the $\vR$ variation found by \cite{Cheng2020}.

Finally, we give each star a relative distance uncertainty of $15\%$, which is slightly larger than the median relative uncertainty of the \cite{BailerJones21} distance estimates. We then make our velocity estimates $\vz$ and $\vphi$ for these mock data.

The results are illustrated in Figure~\ref{fig:FullySmeared}, which shows the angular momentum-vertical velocity space $(\lz,\vz)$, coloured by density in $\vz$ at each $\lz$, for bins between $\ell=130^\circ$ and $230^\circ$. The choice to show the density in $\vz$ rather than in $(\lz,\vz)$ means that we can focus on the changes in $\vz$ with $\lz$, rather than having the structure of the plot being dominated by the decreasing number of stars as a function of $\lz$. These mock observations can be compared to the model distribution, which is shown in Fig.~\ref{fig:Model}. We also show in each panel the typical uncertainty in $L_Z$ and $V_Z$ (defined by the 16th and 84th percentile of the differences between the true values and the derived values). 

\begin{figure*}
	\includegraphics[width=\hsize]{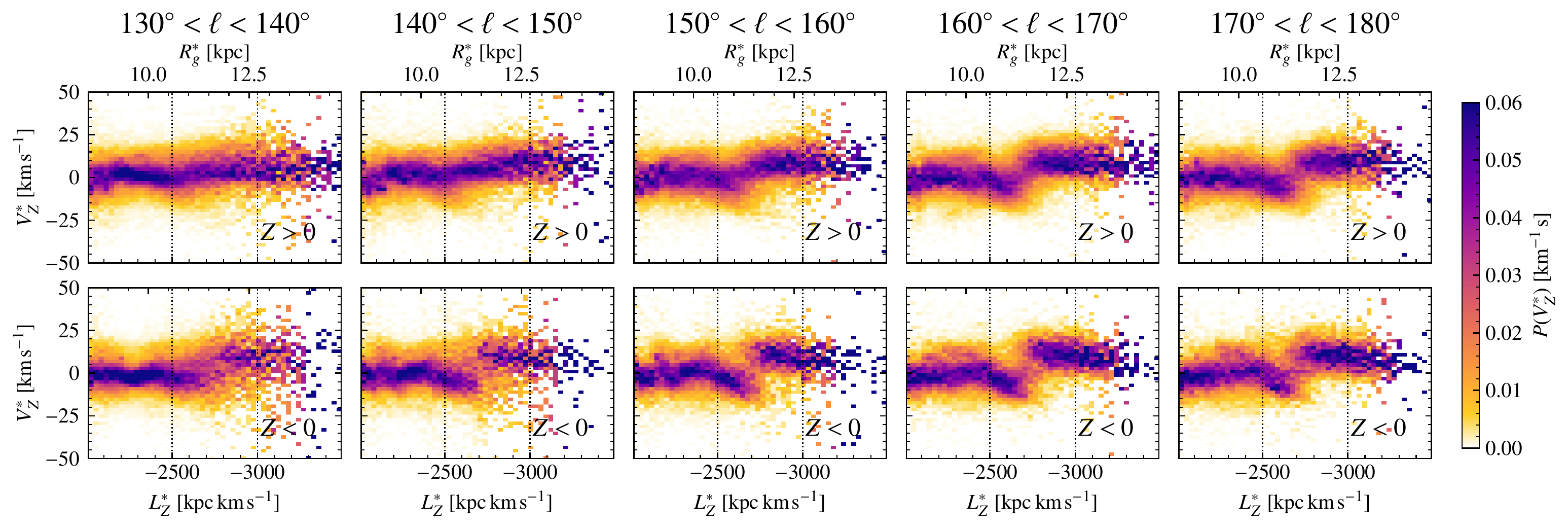}
	\includegraphics[width=\hsize]{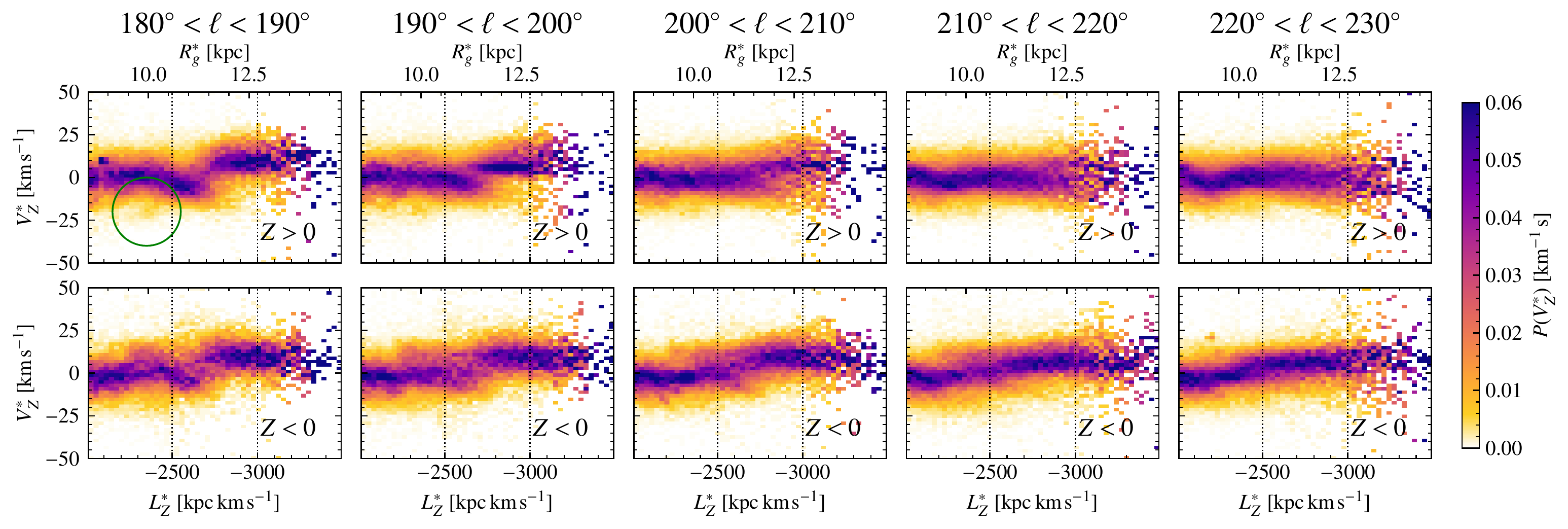}
    \caption{Like Fig~\ref{fig:RgVz}, we show the distribution in $\vz$ of stars in the {\gaia} data as a function of $\lz$ or $\rg$, divided into panels for $130^\circ<\ell<180^\circ$ (upper) $180^\circ<230^\circ$ (lower) and subdivided above and below the plane (as indicated). In this case we compute this only for the young, bright, blue sample of stars indicated in Fig~\ref{fig:HRdiagram}. Again, dotted lines are at $\lz={-}2500$~\&~${-}3000\kms\kpc$ guide the eye. The green circle in the panel for $Z>0$ at $180^\circ< l<190^\circ$ is identical to the one in Fig.~\ref{fig:RgVz}, and the feature it highlights is weaker for this kinematically cool population.
    The `break' at $\lz\approx2750\kms\kpc$ is even more clearly visible than for all stars, and we now see that it extends to $\ell=130^\circ$, but still shows no sign of being present for $\ell>210^\circ$.} 
    \label{fig:RgVzblue}
\end{figure*}

It is obvious from first inspection that the structure in this figure is most clear for bins near the anticentre. Similarly, the measurement uncertainties grow as we move further from the anticentre. This is completely expected given that the approximation described in Sec~\ref{sec:techniques} is insensitive to the real line-of-sight velocity at $\ell=180^\circ$ and becomes increasingly sensitive to it as we move further away across the sky. Another subtler reason for this is that the influence of distance errors becomes greater as we move further from $\ell=180^\circ$, because a given Galactocentric radius $R$ corresponds to an ever increasing distance from the Sun.

Despite the smearing out of the underlying structure, we are still able to see some sign of it out to $50^\circ$ from the anticentre. This informs our understanding when we look at the data -- we should expect perturbations of the galaxy to appear significantly smoothed further from the anticentre, not introduced or accentuated by our approximations. 

In these tests, the distance errors (and the velocity errors they introduce) are the dominant source of blurring of the features of the distribution out to around $\ell=150^\circ$ or $210^\circ$, at which point the uncertainties due to having no line-of-sight velocity measurements begin to dominate. We demonstrate this in Appendix~\ref{app:Mock}, where we isolate the influence of the two effects separately. We also show that the influence of the spiral perturbation is negligible.
Reducing distance uncertainties is almost as important as adding line-of-sight measurements for stars in our efforts to find this structure, even far from the anticentre.


\section{The outer Milky Way}
\label{sec:OuterMilkyWay} 

First, in Figure~\ref{fig:RgVz} we show, for our full sample, angular momentum-vertical velocity space $(\lz,\vz)$, coloured by density in $\vz$ at each $\lz$. This is separated into panels each showing a $10^\circ$ range in Galactic longitude $\ell$, and also split above and below the Galactic plane. The feature found by \citetalias{AnticentrePaper} is again visible in these panels, because in many of the panels there is a downturn in $\vz$ at $\rg\approx10\kpc$ ($\lz\approx2400\kms\kpc$) which goes down to a minimum at $\rg\approx11.5\kpc$ ($\lz\approx2750\kms\kpc$) before what appears to be an abrupt break, with $\vz$ jumping to a positive value.

As we look in bins of increasing $\ell$, this pattern (a downturn followed by an abrupt break) can first be seen below the plane in the $140-150^\circ$ bin (the high $\rg$ part of the $130-140^\circ$ bin is very poorly populated below the plane). The break appears above the plane for the $150-160^\circ$ bin, but remains stronger below the plane up to the $160-170^\circ$ bin. The break is then stronger above the plane for increasing $\ell$, and has faded away below the plane by about $190^\circ$ and above the plane by about $210^\circ$.

This extends our understanding beyond the results of \citetalias{AnticentrePaper}. In that study it was found that the downturn towards lower $\vz$ was dominated by stars below the plane. We see the same for stars in the area of the sky probed by \citetalias{AnticentrePaper} ($170^\circ<\ell<190^\circ$), but can now see that this is not consistent across the sky. 

The panels of Figure~\ref{fig:RgVz} also have another feature worth noting, which is an overdensity near $\vz=-20\kms$ at $\rg\approx10\kpc$, particularly clear for the panels at $Z>0$ for $180^\circ<\ell<200^\circ$, and visible elsewhere. We indicate this feature with a circle in one panel of this figure. This feature is faintly visible in the equivalent plot in \citetalias{AnticentrePaper}, but was not commented upon. This feature is investigated further by Alinder, McMillan \& Bensby (in prep.).

\begin{figure*}
	\includegraphics[width=17cm]{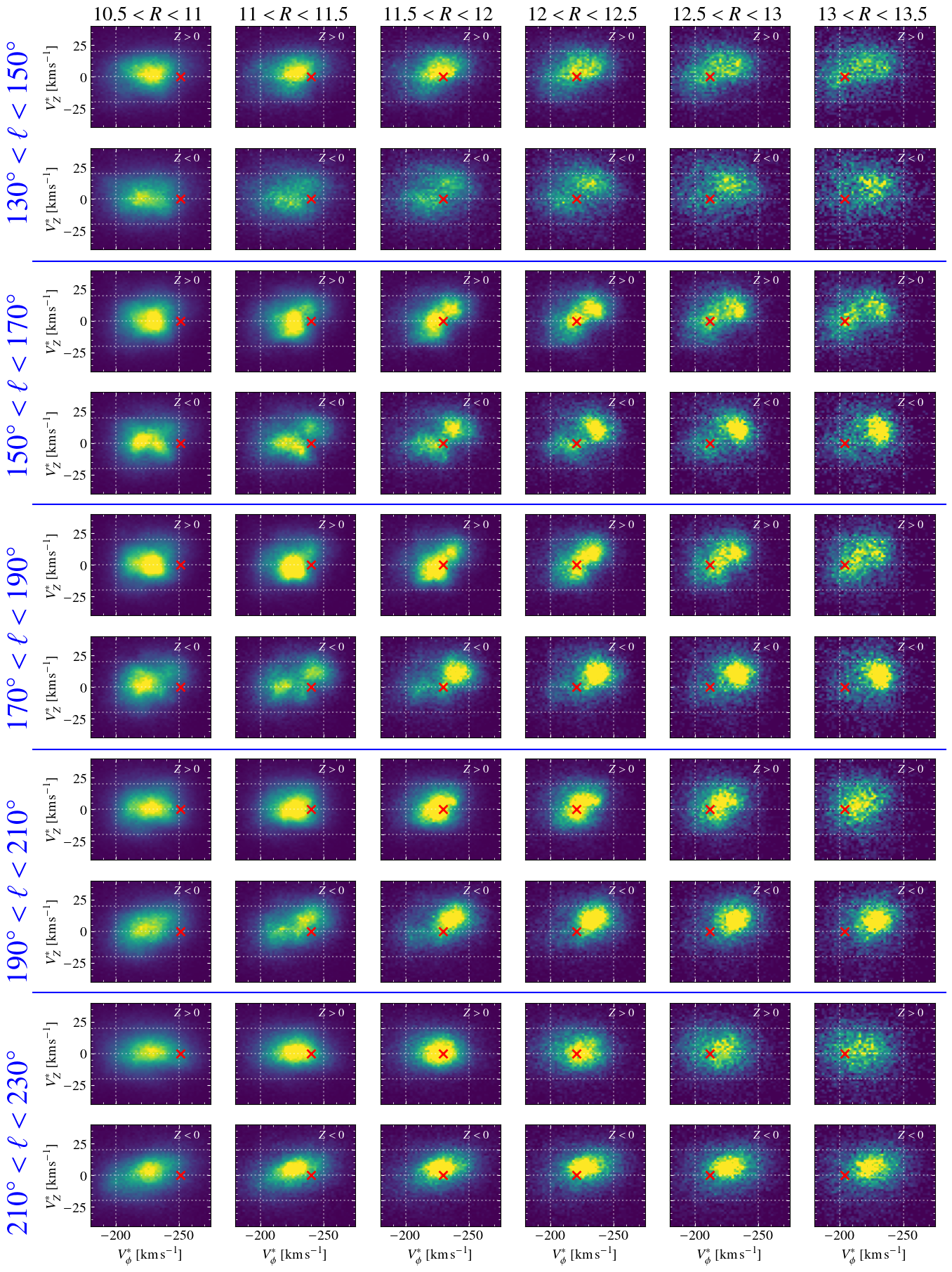}
   \caption{Velocity distributions in the $\vphi{-}\vz$ plane for stars that are divided into bins that are $20^\circ$-wide in $\ell$, $0.5\kpc$ in $R$, and contain stars on one side or another of the Galactic plane (as indicated on the left in blue, at the top, and on each panel, respectively). Dotted grid lines are added to ease comparisons between panels. Bimodal structure is visible in many panels, especially below the plane, with the gap being approximately at $\lz=2750\kms\kpc$, which is marked on each panel with a red cross at $\vz=0$.} 
        \label{fig:velocitydist}
\end{figure*}


\subsection{Young population}
Previous studies have shown that the phase spiral is stronger for stars on kinematically colder orbits, i.e., younger, bluer, stars. \citetalias{AnticentrePaper} also showed that the break at $\rg\approx11.5$ stands out very clearly in their young population (their Fig.~E.8). We therefore show in Figure~\ref{fig:RgVzblue} the $(\lz,\vz)$ plane for the bright young main-sequence sample described in Sect.\ref{sec:data}, again coloured by density in $\vz$. We note also that these are all bright stars and therefore their \gaia\ parallax uncertainties will be relatively small, allowing for more accurate distance estimates even far from the Sun.

The vertical velocity dispersion of these sources is smaller than for the whole population, which is clear in all of these plots. As seen by \citetalias{AnticentrePaper}, this means that the break at $\rg\approx11.5$ stands out clearly, and we have extended this result to a far wider range in $\ell$. The feature near $\vz=-20\kms$ at $\rg\approx10\kpc$ is harder to see for this limited population, presumably because it has a low vertical velocity dispersion so does not sample the phase space around $\vz=-20\kms$ well.


Using these stars as tracers we can see for the first time a hint that the break extends to $\ell=130^\circ$, and therefore perhaps even beyond. The panel for $130^\circ<\ell<140^\circ$ appears to have a break at around $\rg=12\kpc$, but there is no sign of a downtrend before this. There are still no real signs of any break for $\ell>210$, so we tempted to say that it has disappeared. Certainly we can say with some confidence that it is weaker at $\ell=210$ than at $\ell=150$ because, by symmetry, the uncertainties at these two longitudes should be very similar (assuming differences due to differing extinctions are negligible). However, we are very much in the position of having limited data with which to make these judgements, and a more detailed look, and a clearer perspective beyond this range in $\ell$, will have to wait for a larger sample of stars with measured line-of-sight velocities.


\subsection{Velocity distributions}

Finally, in Figure~\ref{fig:velocitydist} we show the velocity distribution in the $\vphi$--$\vz$ plane for $20^\circ$ slices in $\ell$, separated into radial bins of width $0.5\kpc$ from $10.5\kpc$ to $13.5\kpc$, and further divided into stars above and below the plane.

These plots confirm the results seen by \citetalias{AnticentrePaper}, that the break seen in the $\lz-\vz$ plane can equally be thought of as being made by a bimodal velocity distribution, consisting of two clumps of stars in the $\vphi-\vz$ plane, one tending to move downwards with a slower rotational velocity, one moving upwards with a faster rotational velocity. 

To guide the eye, we have put a cross in each panel of these figures at an angular momentum of $2750\kms\kpc$ (for a star at a radius in the middle of the range). The transition we see happens at around this angular momentum, and the gap between the two clumps is at approximately this point in many cases. This confirms that there is a connection between the structure seen here and that seen in Figs~\ref{fig:RgVz}~\&~\ref{fig:RgVzblue}. However it also demonstrates that we lose some information by treating the disturbance as a function of $L_z$, because in some panels the gap in the velocity distribution is at different $\lz$. For example, a number of the panels for $(12.5<R<13)\kpc$ have a substantial clump of stars at the slower rotational velocity and lower $\vz$ that sits almost exactly at $\lz=2750\kms\kpc$. Therefore it is clear that the exact position of the break between the clumps is not simply a value of $L_z$, and the behaviour of the outer Galaxy  in this region can not be completely simplified as in Figs.~\ref{fig:RgVz}~\&~\ref{fig:RgVzblue}.

The break into two clumps can be seen below the plane in all $\ell$ ranges from $130^\circ$ to $210^\circ$, with varying strength and clarity. Above the plane it is only really clear in the range $150^\circ<\ell<190^\circ$.


\begin{figure*}
	\includegraphics[width=\hsize]{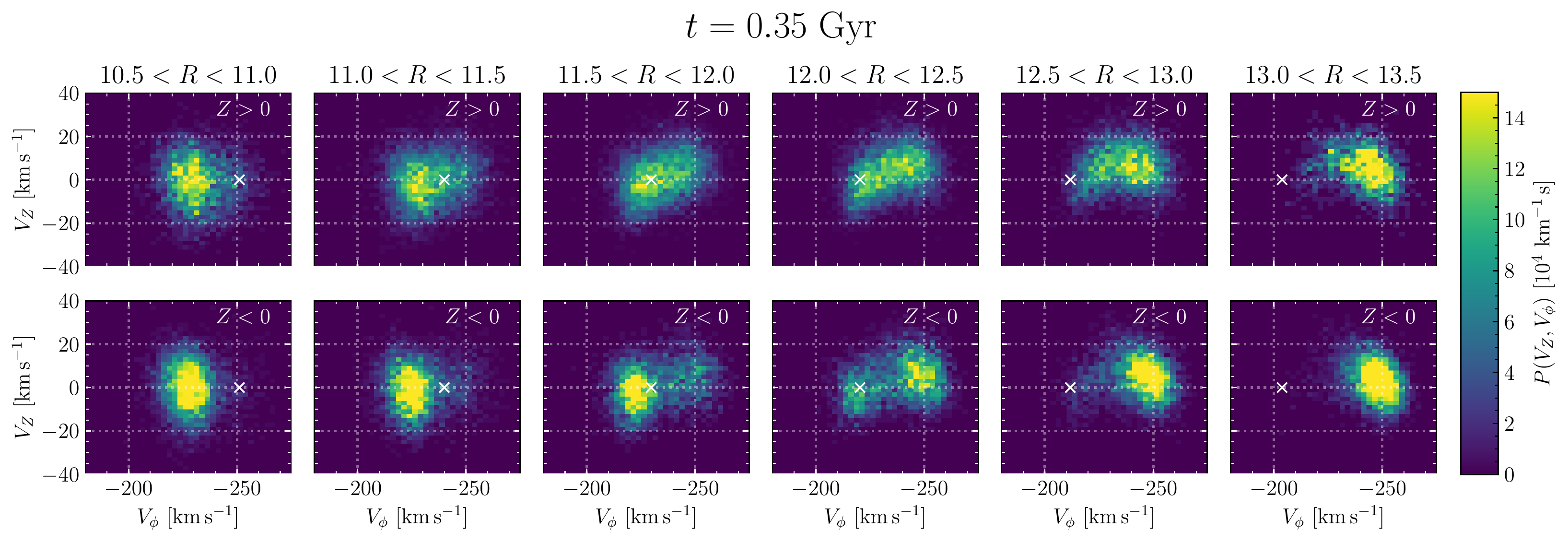}
	\includegraphics[width=\hsize]{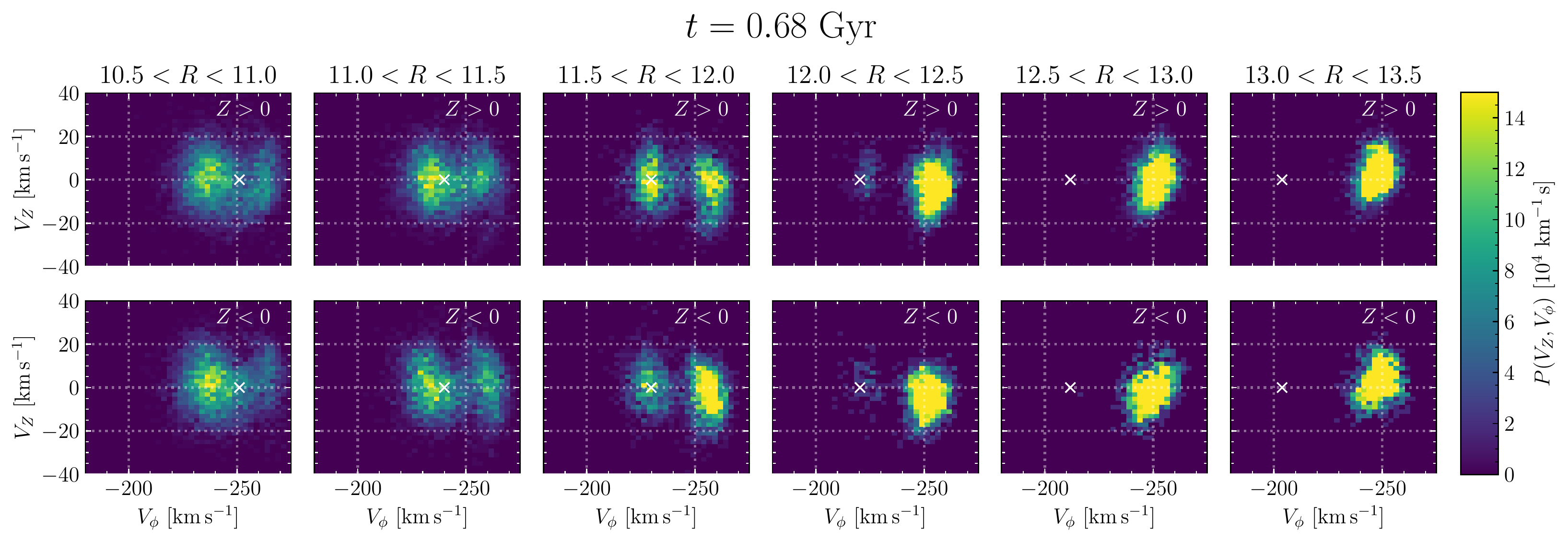}
	\includegraphics[width=\hsize]{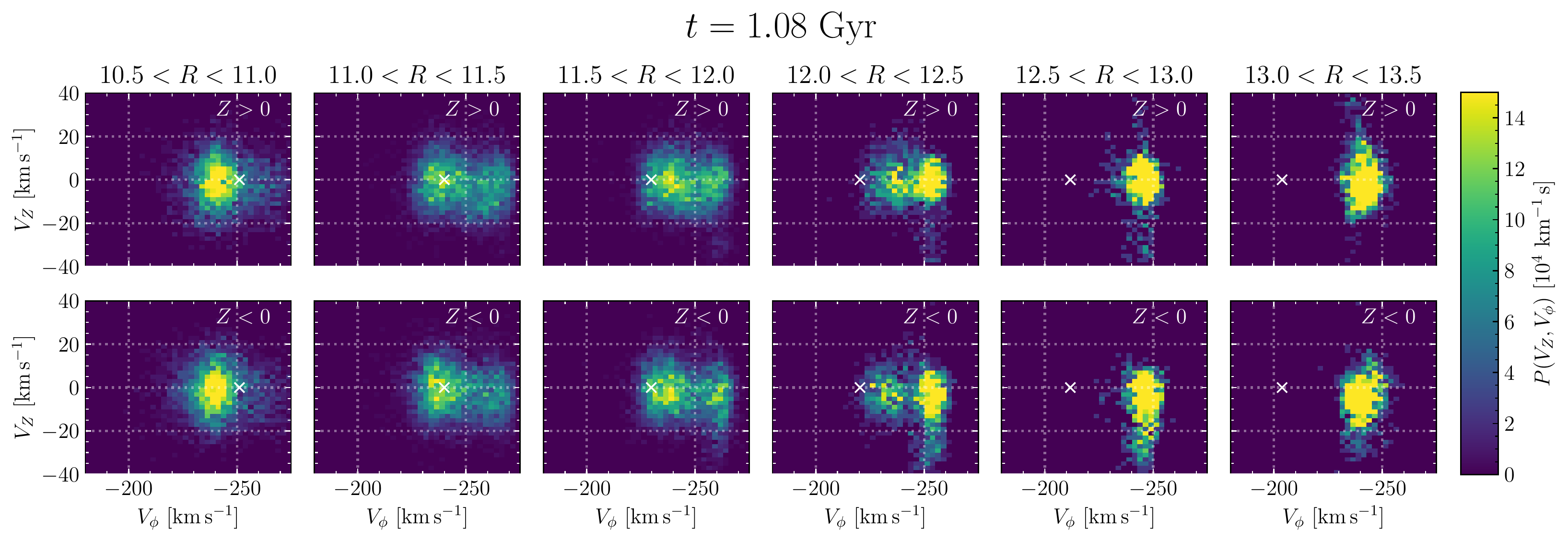}
    \caption{Velocity distributions of a $20^\circ$ segment of the outer disc of the simulated galaxy at three selected times (as indicated, note that the disc crossing of the perturber occurs at $t=100\Myr$). The panels, equivalent to those in Fig~\ref{fig:velocitydist}, show density in the $V_\phi$-$V_Z$ plane for radial bins between $10.5$ and $13.5\kpc$, divided into particles above and below the disc plane. Again, dotted grid lines are added to ease comparisons between panels and with Fig~\ref{fig:velocitydist}. In all these selected times, the velocity distribution is divided into two clumps, but only in the first case are these two clumps displaced from one another in $\vzreal$.} 
    \label{fig:VphiVzSim}
\end{figure*}

\section{Simulations} \label{sec:simulations}

In an effort to interpret what we are seeing in the outer Milky Way, we turn to a numerical simulation of the interaction of an impulsive mass with a cold stellar disc at a single transit point. This is based on the simulations presented by \citet{Bland-Hawthorn2021}, with the only differences being that we have reduced the mass of the perturbing particle from $2\times10^{10} \msun$ to $10^{10} \msun$ and reduced the number of particles. The reduced number of particles allowed us to explore the parameter space of impacts before settling on the simulation we present here, while still allowing us to resolve the behaviour we sought to examine. We found that a $2\times10^{10}\msun$ perturber produced an effect in the outer disc that was significantly larger than that seen in our Milky Way data, while the $10^{10} \msun$ simulation provided a closer match to the perturbation that we see, as the scale of this perturbation scales approximately linearly with the mass of the perturber. We do not claim to have made an exhaustive search of the possible impacts or timings, to have pinned down the mass of the Sgr dwarf when it last had a major impact on the Milky Way disc, or created an exact analogue of the behaviour seen in Sec~\ref{sec:OuterMilkyWay}. That work is reserved for future studies.

\begin{figure*}
	\includegraphics[width=\hsize]{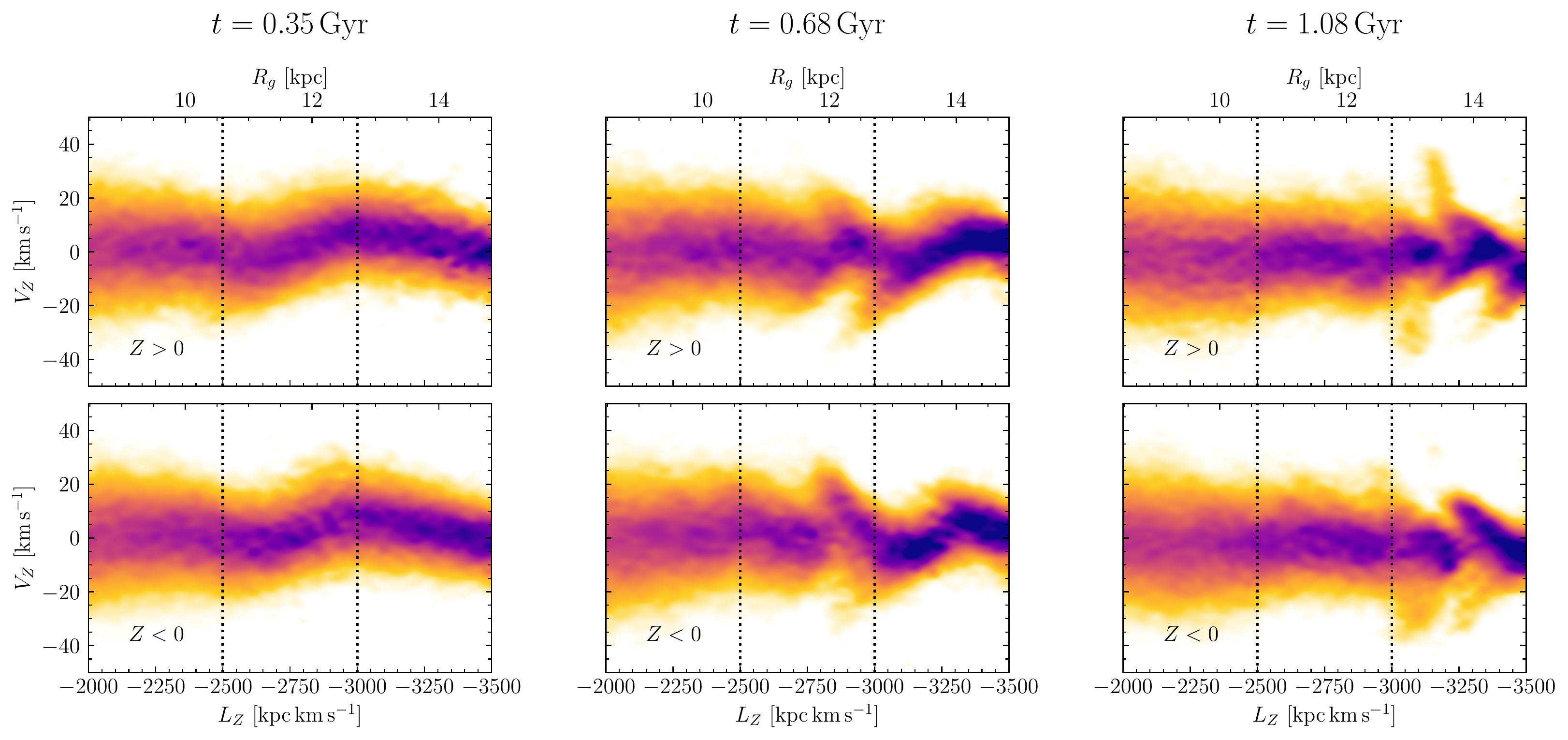}
    \caption{The distribution in $\vzreal$ as a function of  angular momentum $L_z$ in the outer disc of the simulated galaxy at the same three selected times as Fig.~\ref{fig:VphiVzSim}. The panels, equivalent to those in Figs.~\ref{fig:RgVz}~\&~\ref{fig:RgVzblue}, are again 
    divided into the distribution above and below the plane, and we have again drawn dotted lines $\lz={-}2500$~\&~${-}3000\kms\kpc$ to ease the comparison between these equivalent plots.
    We see corrugation of the distribution, which appears over shorter and shorter scales as it winds up over the period shown here, but we do not see a clean break of the kind seen in the Milky Way data.
    } 
    \label{fig:RgVzSim}
\end{figure*}


The interested reader is directed to \cite{Bland-Hawthorn2021} for a full description of this simulation but we will summarise the key elements here. The galaxy model consists of a live $N$-body dark-matter halo, stellar bulge and stellar disc with a total mass of ${\sim}1.45\times10^{12}\msun$ and a nearly flat rotation curve out to $R=20\kpc$. It is constructed from an action-based distribution function using \agama\ such that it is stable in its initial configuration without any relaxation, and its disc is kinematically cold to enhance the signature of perturbations. We populate this model with $10$ million particles in the halo, $5$ million particles in the disc and $2.25$ million particles in the bulge, then evolve it using the $N$-body code \textsc{ramses} \citep{Teyssier2002}. The perturbation is modelled as a nearly impulsive interaction of a point mass which intersects the Galactic plane at $R\approx18\kpc$ travelling with a velocity of ${\sim}330\kms$, and a Galactic azimuth $\phi = 0$.\footnote{\cite{Bland-Hawthorn2021} place the impact and the Sun (at the present day) at $\phi=180^\circ$, and their Galactic disc rotates in the positive sense. We reverse these conventions for consistency with the rest of our analyses.} 
The disc crossing occurs at a simulation time $T\approx100\Myr$, and after $T=150\Myr$ the mass of the perturber begins to exponentially decrease, such that by the time of its second disc crossing it has a mass of $\sim10^{-6}$ times its original mass. This simplifies the encounter and allows us \citep[like][]{Bland-Hawthorn2021} to focus on the effect of a single impact.

\cite{Bland-Hawthorn2021}, following the analytical models of \cite{Binney2018}, showed that the impact of a perturber like this is able to produce a phase spiral in the Solar neighbourhood that is qualitatively similar to that found in Gaia data, and that it will have produced similar phase spirals across the disc (mainly differing in their phase). Here we focus instead on the influence on the outer disc and the effects that we have shown and discussed in Sec.~\ref{sec:OuterMilkyWay}. Specifically, we focus on a circular sector in the galactic disc of width $\Delta \phi=40^\circ$, centred on $\phi=0$. This is roughly the same range in azimuth covered by our Milky Way data 
(see Fig.~\ref{fig:PhiVsl} for a comparison). 

Figure~\ref{fig:VphiVzSim} shows the velocity distribution in this sector for 3 selected snapshots at $t=0.35, 0.68$ and $1.08\Gyr$ in our simulation. Like Fig.~\ref{fig:velocitydist} we show the velocity distribution in radial bins of width $0.5\kpc$ from $10.5\kpc$ to $13.5\kpc$, and show the distribution separately for stars above and below the plane. These snapshots are selected because they are at times when the velocity distribution in this part of the Galaxy separates into two clumps of the kind seen in the Milky Way data in Fig.~\ref{fig:velocitydist}.

Of all of these velocity distributions, it is the one at $0.35\Gyr$ (i.e, $250\Myr$ after the disc crossing of the perturber) that bears the strongest resemblance to the velocity distribution seen in the Milky Way. The two clumps are at different $\vzreal$, with the higher $|\vphireal|$ clump being at positive $\vzreal$ and the lower $|\vphireal|$ clump  at negative $\vzreal$. There is a clear difference between the behaviour above and below the Galactic plane. There is a clearer distinction between two clumps in the velocity space for $Z<0$, and the lower velocity clump has essentially disappeared by the $12.5$-$13.5\kpc$ bin for $Z<0$ while this is not true above the plane.

The other two snapshots do show a velocity distribution divided into two clumps, but they do not have a clear difference in their $\vzreal$ values. We have run this simulation to nearly $4\Gyr$, long after the perturber has reduced to a negligible mass and have found that the velocity distribution continues to  divide into two clumps, then coalesce into a single clump, then divide into two clumps again repeatedly throughout the simulation time. However, the division does becomes weaker, concentrated in the outer bins, and does not show a clear asymmetry either side of $z=0$ or difference in $v_z$ for the two clumps.

To understand this better, in Fig.~\ref{fig:RgVzSim} we look in the simulations at the distribution in $\vzreal$ as it changes with $L_z$, again in the sector of width $\Delta \phi=20^\circ$ around the assumed Solar azimuth in the Galaxy, and in the same times as the panels of Fig.~\ref{fig:VphiVzSim}. In each case there are disturbances in the vertical velocities that are of similar scale to the disturbances seen in the Milky Way. We also see, particularly in the two later snapshots, breaks in the distribution which are comparable to those seen in the Milky Way. The panel at $t=0.35\Gyr$ does not show a clear break in the distribution, but, as we have seen in Fig.~\ref{fig:VphiVzSim}, is the snapshot with a velocity distribution that best matches that seen in the Milky Way.

We would also note that the panels show that the wavelength (in angular momentum) of the disturbance is relatively long shortly after the disc crossing, and gets increasingly short as time passes. This occurs because the disturbance winds up around the disc.

The choice of $\phi$ position to centre this analysis is, to an extent, arbitrary. The initial disc is axisymmetric, and \cite{Bland-Hawthorn2021} used their simulation to explore the behaviour of the phase spiral at a range of positions in the Galaxy. We have explored other choices of $\phi$ and found that the results are, qualitatively, rather similar. The times at which the velocity distribution forms into two clumps varies somewhat with $\phi$, as the disturbance wraps its way around the disc. The general pattern remains the same as for the three snapshots illustrated in Figs~\ref{fig:VphiVzSim}~\&~\ref{fig:RgVzSim}. For the interested reader we have made videos showing the time evolution of the plot shown in Fig.~\ref{fig:VphiVzSim} for choices of $\phi$ at $40^\circ$ intervals available at \url{https://www.astro.lu.se/~paul/MyResearch}.

\section{Discussion} \label{sec:discuss}
\subsection{Galactic seismology}
For over a century, geologists have learned about the interior of the Earth through seismology, studying the waves that pass through it. More recently, helioseismology has revolutionised our understanding of the Sun, and asteroseismology has brought us great insight into the stars of our Galaxy. We have now entered the era of Galactic seismology (or Galactoseismology), with much to learn about our Galaxy from its oscillations.\footnote{For a history of the field and its terminology from \cite{Hunter1963} through \cite{Iye1985} and \cite{Widrow2012} to the present day see Appendix A of \cite{Bland-Hawthorn2021}.}

Galactic seismology differs from the other fields in many ways, of which the two most important are perhaps timescale, and influence on the medium through which the waves are passing. The timescale of the oscillations in the Galaxy are measured in $\Myr$, as opposed to oscillations of the Earth or stars, which can be measured in minutes or hours. We therefore can not wait to follow the oscillations of the Galaxy as they go through their phases. The oscillations of the Earth and the stars can be treated as non-invasive -- not changing the underlying structure of the medium through which they are travelling. This is not true for the oscillations of the Galaxy. \cite{Darling2019} note, following the argument of \cite{HunterToomre1969}, that the influence of the perturbation on the unperturbed disc is of the same order as the influence of the unperturbed disc upon the perturbation. They further show that the influence of self gravity slows the growth of the phase spiral. This result that can also be seen by comparing the phase spiral found by \cite{Bland-Hawthorn2021} with the one seen in the analytic models of \cite{Binney2018} which winds up quickly \citep[see also][]{Hunt2021}.

The bimodality of the velocity distribution in the outer Galaxy discovered by \citetalias{AnticentrePaper} was, like the phase spiral before it, not predicted. Here we have shown that it can be explained, at least qualitatively, by a model that also produces a phase spiral. These different measurements, or phase space projections in which we conduct our analysis, can be thought of as two separate Galactic seismographs, but they must be intrinsically linked and understanding of our Galaxy will come from models that can consistently explain both.

For now we find ourselves like blindfolded people discovering an elephant, each describing what is right in front of them (a trunk, a tail, a leg, etc.) and not being able to agree on what they have found.\footnote{This is a parable that dates back to at least 500 BCE, e.g., \url{https://suttacentral.net/ud6.4}} Similarly we will probably need to simultaneously explain the phase spiral, the velocity bimodality, the Galactic warp, the ridges in $R{-}V_\phi$ space \citep[][\citetalias{AnticentrePaper}]{Kawata2018,Ramos2018}, the variations in $\vR$ as a function of angular momentum \citep[e.g.][]{Friske2019} and Galactic $\phi$ \citep[e.g.][]{Eilers2020}, and even the structures of the outer Galaxy \citep[e.g.,][\citetalias{AnticentrePaper}]{Xu2015,Laporte2022}. This will be complicated by the fact that some of these phenomena, especially those seen in the planar velocities, are affected by the Galactic bar and spiral arms in a way that is likely to be unrelated to the vertical velocity perturbations that we see. Future data from \gaia, or spectroscopic surveys such as WEAVE \citep{Dalton2016} and 4MOST \citep{deJong2019}, will no doubt allow us to identify a new tusk or two, but our task will remain that of piecing all these elements into an over-all picture.

This study has not described the entire elephant, but has extended our understanding of one part of it. We have shown that the velocity bimodality persists across a range of Galactic $\phi$ angle that goes far beyond those probed by \citetalias{AnticentrePaper}, that it changes in character with varying $\phi$, and appears to fade away by $\phi\approx10^\circ$ ($\ell\approx210^\circ$) although, as we have discussed, this is at least partly due to the difficulty in measuring it at these longitudes. The simulations we have performed show that, like the phase spiral, bimodal velocity distributions can be induced by the passage of the Sagittarius dwarf. However, we have not carefully matched the timing of the velocity bimodality in our simulation to that of the formation and evolution of the phase spiral, nor to the past history of Sagittarius. That work is reserved for future study.

\subsection{Density and bending waves}

\cite{Bland-Hawthorn2021} argue that the impact of the Sgr dwarf triggers two different $m=2$ modes in the inner disc: a density wave, and a corrugated bending wave. They associate the phase spiral with the density wave rolling up and down the bending wave. They note in addition that the outer discs' vertical displacement is better described instead as dominated by an $m=1$ bending mode.

The density wave can be expected to create gaps (or simply low density regions) in the angular momentum distribution, which translate into gaps in $\vphireal$ for stars in a given radial bin. The bending wave can mean that the two clumps either side of this gap have different $\vzreal$. This is what we appear to see in Fig.~\ref{fig:VphiVzSim} for the snapshot at $t=0.35\Gyr$, but the snapshots at $t=0.68$ and $1.08\Gyr$ have gaps, which we associate with the density wave, which do not coincide with a significant vertical velocity difference from the bending wave -- the clumps in these two cases have nearly identical vertical velocities. In this picture, the structure seen in the Milky Way can only be produced by a coincidence of a minimum point in the density wave and a point where the gradient of $\vzreal$ in the bending wave is very large.

It is certainly questionable whether this is exactly what we are seeing in the outer Milky Way. The break in the $\lz{-}\vz$ plane seen in Figs.~\ref{fig:RgVz}~\&~\ref{fig:RgVzblue} does appear to be relatively clean at a given $\lz$, but in velocity space (Fig.~\ref{fig:velocitydist}) the divide does not appear to lie along a line of constant $\vphi$, as it does in the simulation (Fig.~\ref{fig:VphiVzSim}). 
Equally, we do not see a clean break in the $L_Z{-}\vzreal$ plane at $t=0.35\Gyr$, as we do in the Milky Way. There is also a very clear difference in the Milky Way between what we see above and below the plane. We do not see this here, but this may be because we are averaging over a wider range in $\phi$.

The timing of the snapshot that seems to best mimic the behaviour in the Milky Way ($t=0.35\Gyr$, i.e., $250\Myr$ after the disc crossing) is also before \cite{Bland-Hawthorn2021} see a phase spiral forming, though it is consistent with the timing of the phase spiral found using a simpler, analytic, approach by \cite{Binney2018}. The later snapshots do have breaks in the $L_Z{-}\vzreal$ plane, but we can clearly discern in each of them that there is an oscillation with an ever decreasing scale length in $L_Z$, which we can associate with the winding up of the bending wave across the Galactic disc. The equivalent distribution in the Milky Way does not appear to have this oscillation on short scales in $L_Z$, instead having a large scale variation with an abrupt break. We also use a perturber that has a mass which half that of the \cite{Bland-Hawthorn2021} simulation that reproduced the phase spiral, in order to match the scale of perturbation seen in the outer Milky Way, so cannot claim that we simultaneously model both.

Certainly we have only begun to interpret the velocity bimodality in the outer Milky Way disc. One can hope that this will prove a fruitful area, both for learning about the mass and timing of the most recent impacts on the Milky Way disc, and for measuring the properties of the disc too. If the bimodality can indeed only be explain by a coincidence of the density and bending wave then this sets a strong constraint on the impact's timing relative to orbital frequencies in the disc. We might reasonably expect that in the outer disc, as the disc itself becomes less dominant, the influence of self-gravity becomes less important, allowing us to make a unambiguous estimate of these properties and to use simpler, faster, modelling techniques of the kind employed by \cite{Binney2018}.


\section{Summary} \label{sec:conclusions}

In this paper we have shown that the velocity bimodality -- or, equivalently, break in the $L_Z{-}\vzreal$ distribution -- found in the Galactic anticentre by \citetalias{AnticentrePaper} persists across some significant fraction of the Galactic disc. It can be found out to $\ell=130^\circ$, which corresponds to a Galactocentric angle of $\phi\approx20^\circ$, but appears to be fading away by $\ell=210^\circ$ ($\phi\approx10^\circ$). Its behaviour is, as seen by \citetalias{AnticentrePaper}, different above and below the Galactic plane, with a stronger break/bimodality usually visible below the plane. We note that selecting young main-sequence stars allows us to see a stronger effect, which is consistent with previous results that show younger, kinematically colder, stars are more affected by disturbances in the vertical velocities of stars \citep[e.g.,][]{Bland-Hawthorn2019}.

Using rather idealized $N$-body simulations we have shown that the passage of a Sagittarius-like dwarf galaxy is capable of producing qualitatively similar effects in a Galactic disc. These simulations are also capable of producing a phase spiral akin to the one which has been discovered in the Solar neighbourhood.

This study has been conducted entirely without measured line-of-sight velocities for the stars considered. For the foreseeable future we will be in the position that we have measured proper motions for far more stars than line-of-sight velocities. This is due to both the two billion stars observed by Gaia, and the still larger number which would be observed by a successor mission conducting astrometry in the infrared \citep{Hobbs2021}. Exploiting these incomplete 5D phase-space data are absolutely key to building up large samples of stars for study, which allows us to either constrain our models far more accurately or, as in the case of the velocity bimodality, find features of our Galaxy that we have not predicted using our models.
Our investigation of the uncertainties inherent in our approach makes the further point that distance uncertainties are as serious a barrier to characterising sharp structure in velocity space as the absence of line-of-sight velocity measurements.


\section*{Software}
This study utilised the Python packages 
\texttt{numpy} \citep{numpy}, 
\texttt{matplotlib} \citep{matplotlib}, 
\texttt{pandas} \citep{pandas} and 
\texttt{dustmaps} \citep{2018JOSS....3..695M}.

\section*{Acknowledgements}
PM and JP gratefully acknowledge support from project grants from the 
Swedish Research Council (Vetenskapr\aa det, Reg: 2017-03721; 2021-04153). Some of the computations in this project were completed on computing equipment bought with a grant from The Royal Physiographic Society in Lund.
TTG acknowledges partial financial support from the Australian Research Council (ARC) through an Australian Laureate Fellowship awarded to JBH.
LC acknowledges financial support from ANID/Fondecyt Regular Project 1210992. 
TA acknowledges the grant RYC2018-025968-I funded by MCIN/AEI/10.13039/501100011033 and by ``ESF Investing in your future''.
TA, MRG and FF acknowledge that this work was partially funded by the Spanish MICIN/AEI/10.13039/501100011033 and by ``ERDF A way of making Europe'' by the ``European Union'' through grant RTI2018-095076-B-C21, and the Institute of Cosmos Sciences University of Barcelona (ICCUB, Unidad de Excelencia `María de Maeztu') through grant CEX2019-000918-M.
PR acknowledges funding from the University of Barcelona, Margarita Salas grant (NextGenerationEU).
This work has made use of data from the European Space Agency (ESA) mission
{\it Gaia} 
(\url{https://www.cosmos.esa.int/gaia}), processed by the {\it Gaia}
Data Processing and Analysis Consortium (DPAC,
\url{https://www.cosmos.esa.int/web/gaia/dpac/consortium}). Funding for the DPAC
has been provided by national institutions, in particular the institutions
participating in the {\it Gaia} Multilateral Agreement.

\section*{Data availability}
The observational data underlying this article were accessed from the \gaia\ archive (\url{https://gea.esac.esa.int/archive/}). The derived data generated in this research,  mock data used in Sec.~\ref{sec:CheckvR}, and simulation data from Sec.~\ref{sec:simulations} will be shared on reasonable request to the corresponding author.



\bibliographystyle{mnras}
\bibliography{OuterGalaxy} 



\appendix

\section{Derivation of the approximations} \label{app:Convert}

Taking the usual convention for Heliocentric coordinates, that $U$ is the component of velocity towards $(l,b)=(0,0)$, $V$ the component towards $(l,b)=(90^\circ,0)$ and $W$ towards $(l,b)=(0,90^\circ)$, we can find the components of Heliocentric velocity \emph{excluding line-of-sight velocity} as 
\begin{multline}
     (U^\bot, V^\bot, W^\bot) = (-\vl \sin l - \vb \cos l \sin b, \\ 
        \vl \cos l -  \vb \sin l\sin b, \; \vb\cos b)
\end{multline}

where the missing line-of-sight velocity decomposes into 
$(U^\parallel,V^\parallel,W^\parallel)= v_\parallel(\cos l\cos b,\sin l\cos b,\sin b)$, and $U = U^\bot+U^\parallel$ etc.

We can convert these into Galactocentric coordinates using the values for the Sun's position and velocity given in Section \ref{sec:data}. We place the Sun at $(X,Y,Z) = (\rsun,0,\zsun)$, meaning that we have to flip the orientation of the velocity axes in the plane from that of our Heliocentric system. We have  
\begin{equation}
    (v_X^\bot,v_Y^\bot,v_Z^\bot) = (-(U^\bot+\usun),-(V^\bot+\vsun),W^\bot+\wsun)
\end{equation}
To make further progress we have to introduce our approximation that $\vR=0$. From the above, we can see that 
\begin{equation} \label{eq:app:vR}
    \vR = (v_X^\bot-\vlos\cos l\cos b) \frac{X}{R} + (v_Y^\bot-\vlos\sin l\cos b) \frac{Y}{R}
\end{equation}
which, under the approximation that $\vR=0$, can be rearranged to give an estimate of $\vlos$, which we refer to as $\vlos^*$ where
\begin{equation}
    \vlos^* = \frac{X v_X^\bot +Y v_Y^\bot}{(X\cos l + Y\sin l)\cos b}
\end{equation}
This allows us to find our estimates 
\begin{align}
 \vphi &= \frac{X(v_Y^\bot-\vlos^*\sin l\cos b) -Y(v_X^\bot-\vlos^*\cos l\cos b)}{R}\nonumber \\
 & =\frac{R(-v_X^\bot\sin l + v_Y^\bot\cos l)}{X\cos l + Y \sin l}\label{eq:app:vphi},
\end{align}
\begin{equation}
    \vz = v_Z^\bot + \vlos^* \sin b.
\end{equation}

We can find the error associated with a star having a true $\vR$ by placing this in eq.~\ref{eq:app:vR} and following the equations through. We find errors $\epsilon_{\phi} = \vphireal-\vphi$ and $\epsilon_{Z} = \vzreal-\vz$ to be
\begin{equation}
    \epsilon_{\phi} = \vR \; \frac{X\sin l - Y\cos l}{X\cos l + Y \sin l}
\end{equation}
and
\begin{equation}
    \epsilon_{Z} = - \vR \tan b \; \frac{R}{X\cos l + Y\sin l}.
\end{equation}
 $\epsilon_{\phi}$ runs from ${\sim}-3\vR$ at $\ell=130^\circ$ to ${\sim}3\vR$ at $\ell=230^\circ$, with a mild dependence on distance (in the sense that more distant stars are worse affected). $\epsilon_{Z} = \vR\tan b$ at $\ell=180^\circ$, increasing symmetrically about $\ell=180^\circ$ to ${\sim}3\vR\tan b$ at $\ell=130^\circ$ and $\ell=230^\circ$, again with somewhat greater errors at greater distances. We note that $|\tan b|<0.18$ for all stars in our sample ($|b|<10^\circ$). 

\section{Details of the mock data} \label{app:Mock}

\begin{table*}
 	\centering
 	\caption{Parameters of the distribution functions used to sample the mock data used to validate our assumptions in Sec.~\ref{sec:CheckvR}. The parameters have the meanings described in App.~\ref{app:Mock}.}
 	\label{tab:df}
\begin{tabular}{lcccccccc}
\hline
Component & $\Sigma_0\,[\msun\kpc^{-2}]$ & $R_{\rm disc}\,$[$\kpc$] & $h\,$[$\kpc$] & $\sigma_{R,0}\, $[$\kms$] 
& $R_{\sigma_R}\, $[$\kpc$] & $\beta$ & $\tau_{\rm SFR}\,[\tau_{\rm max}]$& $\xi$ \\ 
\hline
Thin disc & $8.9\times10^{8}$ & $2.5$ & $0.3$ & $90$ &  $15$ & $0.33$ & $0.8$ & $0.225$ \\ 
Thick disc & $1.8\times10^8$ & $3.0$ & $0.7$ & $180$ &  $6$ & - & - & - \\
\hline
- & $M\,[\msun]$ & $J_0\,[\kms\mathrm{kpc}]$ & $\Gamma$ & $\beta$ & $h_r$ & $h_z$  &  $J_{\rm max}\,[\kms\kpc]$ & $\zeta$  \\
\hline
Halo & $1.5\times10^{10}$ & $500$ & $0$ & $3.5$ & $1.6$ & $0.7$  & $1\times10^5$ & $2$  \\
 		\hline
 	\end{tabular}
 \end{table*}
 
\begin{figure}
\centering
	\includegraphics[width=0.7\hsize]{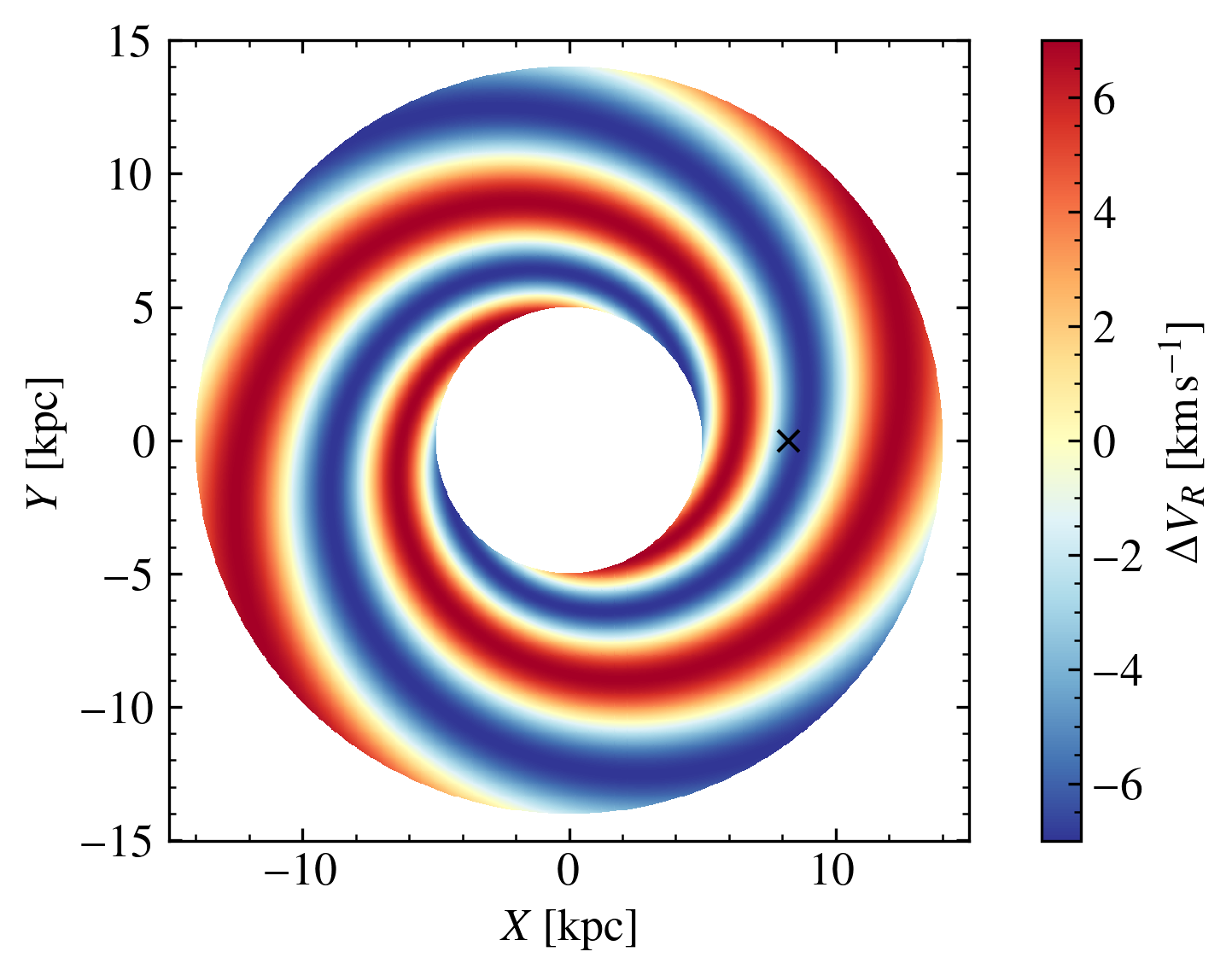}
    \caption{Velocity perturbation in $\vR$ applied to our mock data. The perturbation is comparable to that seen in the Milky Way, but somewhat larger, to provide a more stringent test.} 
    \label{fig:Spiral}
\end{figure}

The dynamical model on which we base the experiments in section~\ref{sec:CheckvR} has two discs and a halo, each of which is defined such that it is in equilibrium within the axisymmetric Milky Way gravitational potential taken from \cite{McMillan17}. We use the \textit{St\"ackel fudge} approach of \cite{Binney2012} to derive the actions $\boldsymbol{J}=(J_r,J_z,J_\phi\equiv L_Z)$ as a function of position and velocity.

\begin{figure*}
	\includegraphics[width=\hsize]{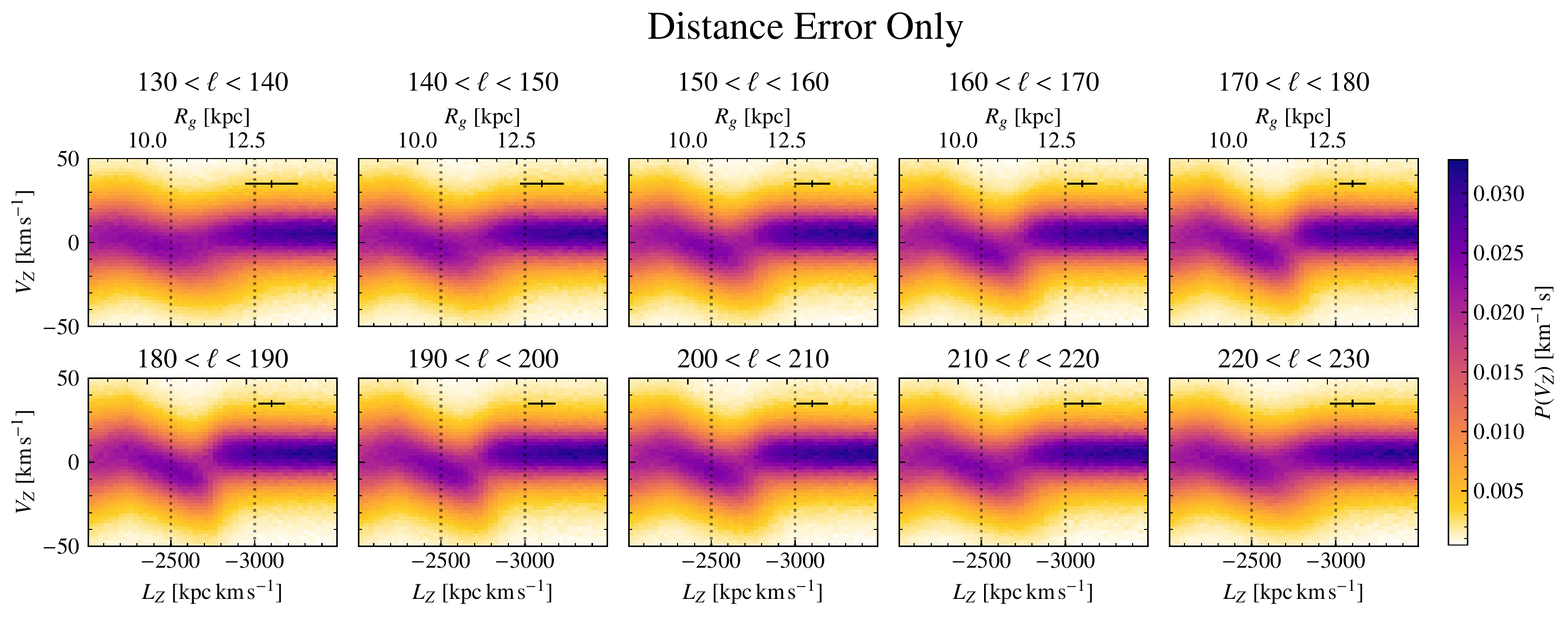}
    \caption{Density in $\vzreal$ at different $L_z$ for mock data with realistic distance uncertainties \emph{but with precise line-of-sight velocities and no spiral perturbation}. As with Fig.~\ref{fig:FullySmeared}, each panel shows the results for different $10^\circ$ ranges in $\ell$. Error bars indicate the typical uncertainties in $L_z$ and $\vzreal$. The error introduced is a significant fraction of the total uncertainty seen in Fig.~\ref{fig:FullySmeared} in all panels, and is the dominant source of uncertainty near the anticentre.}
    \label{fig:DistanceSmeared}
\end{figure*}

\begin{figure*}
	\includegraphics[width=\hsize]{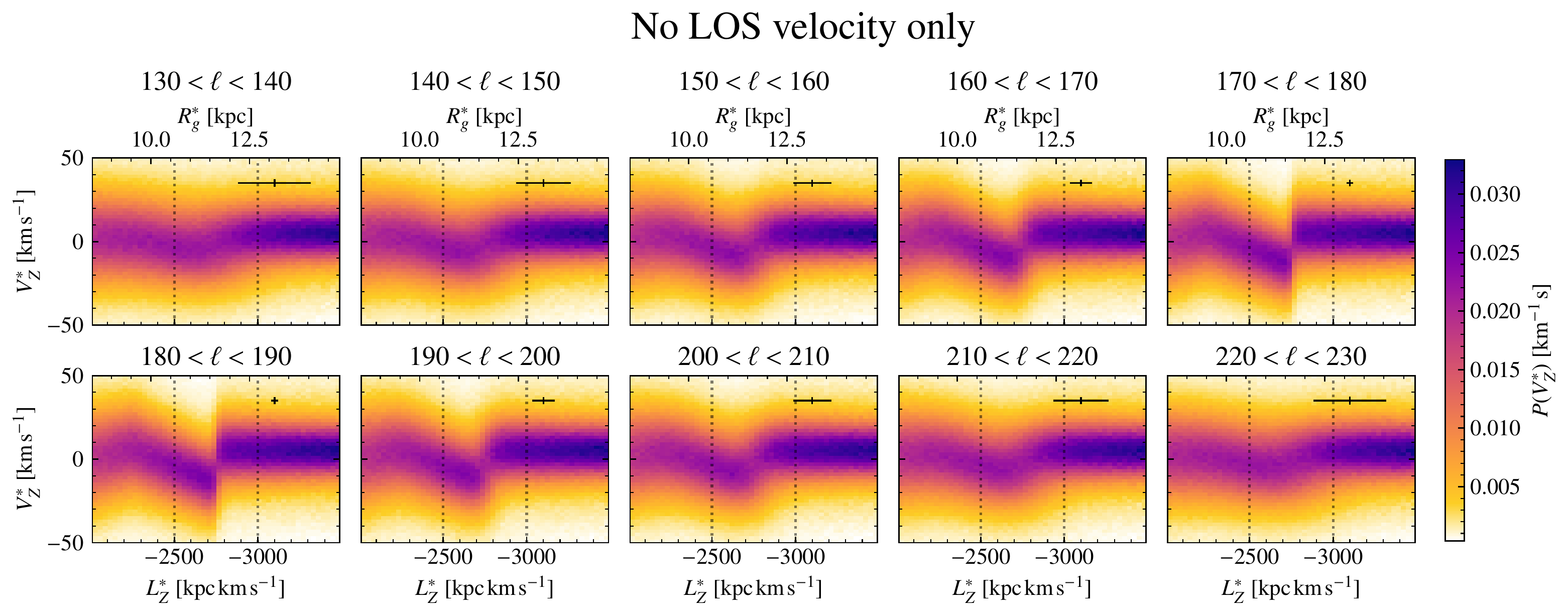}
    \caption{Density in $\vz$ at different $\lz$ for mock data with \emph{no distance errors or spiral perturbation}, but with no line-of-sight velocity measurement. Each panel shows the results for different $10^\circ$ ranges in $\ell$. The error introduced by the lack of line-of-sight velocities is never much larger than that introduced by the distance uncertainties (Fig~\ref{fig:DistanceSmeared}), and only becomes the dominant source of uncertainty ${\sim}30^\circ$ from the anticentre.} 
    \label{fig:VlosSmeared}
\end{figure*}

\begin{figure*}
	\includegraphics[width=\hsize]{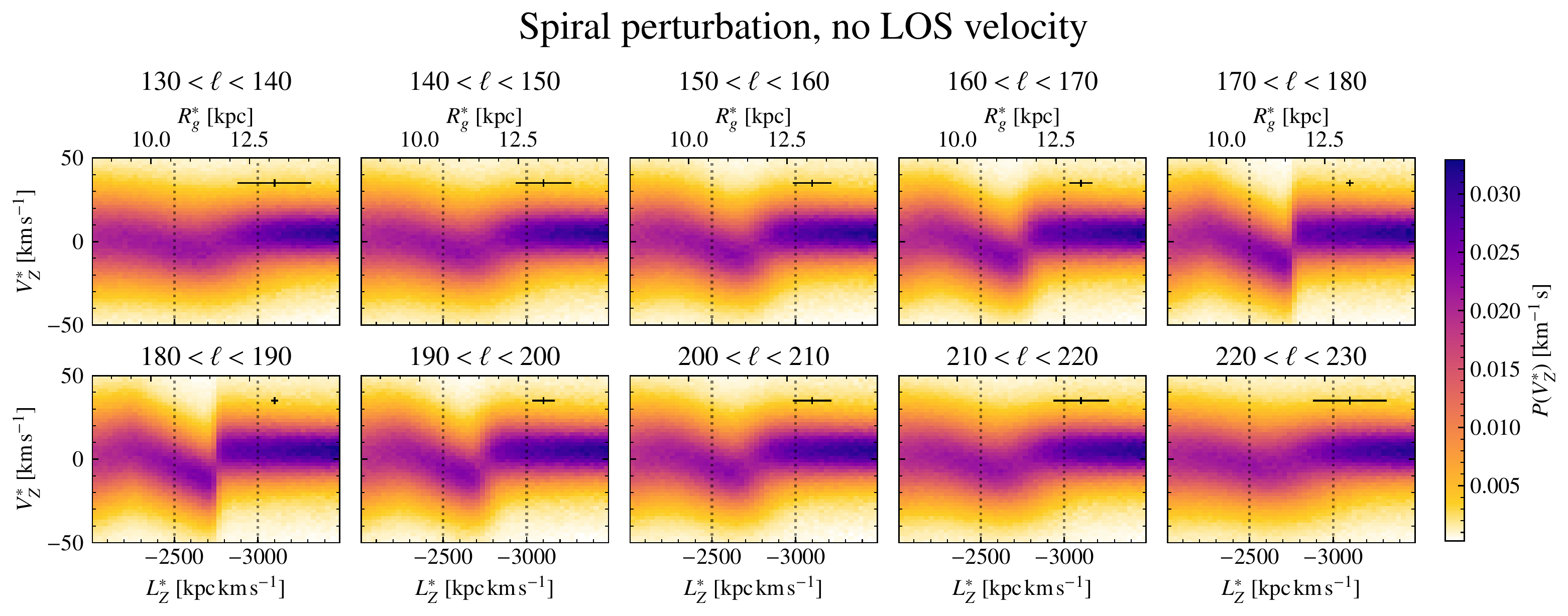}
    \caption{Density in $\vz$ at different $\lz$ for mock data with \emph{no distance uncertainties}, but with a spiral perturbation and no line-of-sight velocity measurement (i.e., differing from Fig~\ref{fig:VlosSmeared} only because of the re-introduction of the spiral perturbation). The results are almost indistinguishable from those in Fig~\ref{fig:VlosSmeared}, illustrating the very limited effect of the spiral perturbation on these results.} 
    \label{fig:VlosSpiralSmeared}
\end{figure*}

The discs are based on the quasi-isothermal distribution function \citep{BinneyMcMillan2011} as modified in \agama,
\begin{align}  \label{eq:DFdisk_quasiiso}
f(\boldsymbol{J}) = &\frac{\Sigma_0\,\Omega(R_\circ)}{2\pi^2\,\kappa(R_\circ)^2}
\exp\bigg(-\frac{R_\circ}{R_\mathrm{disc}}\bigg) \times f_\pm(J_\phi) \\
&\times \frac{\kappa(R_\circ)}{\tilde\sigma_R^2(R_\circ)}
\exp\bigg(-\frac{\kappa(R_\circ)\,J_r}{\tilde\sigma_R^2(R_\circ)}\bigg) \nonumber\\
&\times \frac{\nu(R_\circ)}{\tilde\sigma_z^2(R_\circ)}
\exp\bigg(-\frac{\nu(R_\circ)\,J_z}{\tilde\sigma_z^2(R_\circ)}\bigg), \nonumber
\end{align}
 with $(\kappa,\nu,\Omega)$ being the epicycle frequencies. $R_\circ$ is an estimate of the typical orbital radius, which is given by radius of a circular orbit with angular momentum $L_{\rm circ}=L_z+J_r+J_z/4$. The surface density of this disc is $\Sigma\approx\Sigma_0\exp(-R/R_\mathrm{disc})$, and the functions $\tilde\sigma_R$ and $\tilde\sigma_z$, which approximately give the velocity dispersions in $R$ and $z$ (and therefore in the latter case also determine the disc scale height), are of the form
\begin{align}
    \tilde\sigma^2_z(R_\circ) &= 2\,h^2\,\nu^2(R_\circ) + \sigma^2_{\rm min} \label{eq:sigmas}\\
    \tilde\sigma^2_R(R_\circ) &= \sigma^2_{R,0}\, \exp(-2R_\circ / R_{\sigma_R}) + \sigma^2_{\rm min}\nonumber,
\end{align}
where $\sigma_{\rm min} = 5 \kms$ is introduced to avoid numerical difficulties at large $R_\circ$. The term $f_\pm(J_\phi)$ is a function which is much larger for negative $J_\phi$ than the equivalent positive $J_\phi$. This gives the model a negative sense of rotation, like the Milky Way. Omitting it would give the distribution function of a disc that has equal numbers of stars rotating in opposite directions.

The thick disc distribution function is taken to be a single quasi-isothermal distribution, while the thin disc one is a superposition of quasi-isothermal populations taken to be of ages $\tau$, weighted proportional to $\exp({\tau/\tau_{\rm SFR}})$ in order to mimic the thin disc's expected decreasing star formation rate. The radial and vertical dispersions of the quasi-isothermal populations both vary to reflect the age-velocity dispersion relation in the form $\sigma(\tau) = \sigma_1 \big[\tau + (1-\tau)\xi^{1/\beta} \big]^\beta$, with $\beta\simeq 0.33$ being the power law associated with the age-velocity dispersion relation, and $\xi$ being the ratio of the velocity dispersion of the oldest population and the youngest population.

The stellar halo is based on the double power-law form introduced by \cite{Posti2015}, as modified in \agama
\begin{align}  \label{eq:DFhalo}
f(\boldsymbol{J}) &= \frac{M}{(2\pi\, J_0)^3} 
\left[1 + \left(\frac{J_0}{h(\boldsymbol{J})}\right)^\eta \right]^{\Gamma/\eta}
\left[1 + \left(\frac{g(\boldsymbol{J})}{J_0}\right)^\eta \right]^{\frac{\Gamma-\mathrm{B}}{\eta}} \nonumber \\
&\times \exp\bigg[-\left(\frac{g(\boldsymbol{J})}{J_\mathrm{max}}\right)^\zeta\bigg],
\end{align}
where
\begin{align*}
g(\boldsymbol{J}) &\equiv  J_r + J_z + |J_\phi|, \nonumber\\ 
h(\boldsymbol{J}) &\equiv h_r J_r + h_z J_z   + h_\phi   |J_\phi|  \nonumber
\end{align*}

The interested reader is directed to \cite{Vasiliev2018,Vasiliev2019} for more details. Parameters of the distribution functions of all three components are given in  table~\ref{tab:df}.

The samples drawn from this model are perturbed in the way described in~\ref{sec:CheckvR}, which produces the disturbance in $\vzreal$ illustrated in figure~\ref{fig:Model} and a disturbance in $\vR$ which is illustrated in Fig~\ref{fig:Spiral}, and is broadly similar to that found by \cite{Eilers2020}.

\subsection{Different sources of error}

Figures~\ref{fig:DistanceSmeared},~\ref{fig:VlosSmeared}~\&~\ref{fig:VlosSpiralSmeared} subdivide the effects which blur out the structure in the $L_Z-\vzreal$ plane. Figure~\ref{fig:DistanceSmeared} shows the effect of the 15 percent distance uncertainty applied to the model on its own (i.e., assuming that line-of-sight velocities are known). This causes blurring in all panels that increases gradually as we move further from the anticentre, as the distance we have to look to see stars at a given $R$ (which generally have $R_g{\approx}R$) grows with distance from the anticentre. Note that we give quantities as, for example, $L_Z$ rather than $\lz$ because we are not having to estimate these quantities without line-of-sight velocities.

Figure~\ref{fig:VlosSmeared} shows the effect of assuming that there is no measured line-of-sight velocity, even if the distance to the star is known precisely. The uncertainty introduced is very small towards the anticentre -- far smaller than the uncertainty introduced by the distance uncertainty. However, the uncertainty introduced by this approximation grows with increasing angle from the anticentre faster than the uncertainty associated with the distance, and is larger by $\ell\approx150^\circ$. Figure~\ref{fig:VlosSpiralSmeared} show that the effect of the spiral in addition to the lack of line-of-sight velocity is so small as to be completely negligible. We do not attempt to isolate the effect of the spiral completely because, if the line-of-sight velocity is known, we no longer have to to make the approximation that $\vR=0$, so the spiral perturbation has no effect on our results. 

\bsp	
\label{lastpage}
\end{document}